\documentclass[letter,nofootinbib,preprintnumbers,amsmath,amssymb,floatfix]{revtex4}

\usepackage{graphicx}
\usepackage{dcolumn}
\usepackage{bm}
\usepackage{slashed}
\usepackage{amssymb}
\usepackage{color}
\usepackage{subfigure}

\newcommand{\beqa}{\begin{eqnarray}}
\newcommand{\eeqa}{\end{eqnarray}}
\newcommand{\be}{\begin{equation}}
\newcommand{\ee}{\end{equation}}
\newcommand{\nn}{\nonumber\\}
\def\Tr{\mbox{Tr}}
\def\Im{\mbox{Im}}

\def\st{\begin{equation}}
\def\stp{\end{equation}}
\def\V{{\mathcal V}}

\def\Eq#1{Eq.~(\ref{#1})}
\def\x{{\bm x}}
\newcommand{\llangle}{\left\langle}
\newcommand{\rrangle}{\right\rangle}

\begin{document}

\title{Direct Photon Elliptic Flow at BNL Relativistic Heavy Ion Collider (RHIC) and CERN Large Hadron Collider (LHC)}

\author{Young-Min Kim,$^a$ Chang-Hwan Lee,$^a$ Derek Teaney,$^b$ and Ismail Zahed$^b$}

\affiliation{a) Department of Physics, Pusan National University, Busan 46241, Rep. of Korea\\
b) Department of Physics and Astronomy, Stony Brook University, NY 11794, USA}

\date{\today}

\begin{abstract}
We use an event-by-event hydrodynamical description of the heavy-ion collision process with Glauber initial
conditions to calculate the thermal emission of
photons. The photon rates in the hadronic phase follow from a
spectral function approach and a density expansion, while in the partonic phase they follow
from the AMY perturbative rates.
The calculated photon elliptic flows are lower than
those reported recently by both the ALICE and PHENIX collaborations.
\end{abstract}

\maketitle

\section{Introduction}

A chief finding of the heavy ion program both at RHIC and LHC is a new state of  matter under extreme conditions, the strongly coupled quark gluon plasma (sQGP)~\cite{Heinz:2013th,Luzum:2013yya,Teaney:2009qa}. The prompt
release of a large entropy in the early partonic phase together with a rapid thermalization and 
short mean free paths, suggest a nearly perfect fluid with a shear viscosity
almost at its quantum bound. The time evolution of the fluid follows the laws of relativistic 
hydrodynamics. Detailed analyses of the hadronic spectra, including their $p_T$ distributions
and azimuthal anisotropies have put some reliable constraints on the main  characteristics
of the relativistic viscous fluid, namely its  shear viscosity. 

The electromagnetic emissions in relativistic heavy ion collisions are thermal  at low and intermediate 
$p_T$ \cite{Lee14,Gal15,RALF}. They are dominated by perturbative processes at high $p_T$. 
Since the photons interact very weakly on their way out, they are ideal
for a better understanding of the hadronic composition, evolution and spatial anisotropies of this fluid.
They provide additional constraints on our understanding of the sQGP. Detailed analyses of the 
photon emissivities both from the partonic and hadronic phases, have led to some  understanding
of the overall photon yield at low and intermediate mass~\cite{Luzum:2013yya}. It is the purpose of this paper to extend
these analyses and results to the recently reported anisotropies at both colliders~\cite{Adl03,PHE12,PHE16,ALICE12,ALICE13,CMS13,CMS14}.

To follow the evolution of the fluid, we will use an improved hydrodynamical model developed by one of 
us~\cite{Mazeliauskas:2015vea}. On an event-by-event basis, the  model is initialized using the Glauber model, and its parameters
are constrained by the measured charged multiplicities for fixed centralities~\cite{Adl04,Adl03,ALICE13,CMS13,ALICE}. 
The model yields reasonable event-by-event hadronic elliptic flows in semi-central collisions at both collider 
energies. We will use it to critically examine the photon anisotropies emanating from the partonic and
hadronic composition of this hydrodynamical model.

Our analysis complements a number of 
recent theoretical studies of these anisotropies~\cite{GALE,Gal15,RALF}, although 
the analysis is,  to a certain extent, less complete.  Relative to other works, one 
of the major differences is the hadronic photon production rate, which 
were revisited by two of the authors~\cite{Lee14}. We have 
not included the first viscous correction when computing the hadronic
or partonic photon production rates. These corrections should ultimately
be included in our computation. Nevertheless, we believe that the current
comparison to data is complete enough to be of considerable value. First,
comparing fairly to data requires a sizable code base, involving event-by-event viscous hydrodynamics and tabulated rates. It is certainly a good idea 
for more than one group to undertake such calculations. Looking forward
to the Beam Energy Scan at the RHIC collider, the current calculation 
develops the computational machinery to compute both photons and dileptons
at lower energies where the hadronic rates play an increasingly important
role.

In section~\ref{sec2}, we briefly review the physical content of both the hadronic and partonic rates to be used in this analysis. In section~\ref{secani} we define the various azimuthal moments of the photon emission rates 
both in transverse momentum and rapidity. In section~\ref{sechydro} we briefly overview the hydrodynamical 
set up for the space-time evolution of the fireball using the Glauber model for initial conditions. 
In section~\ref{sec_pp}, we summarize our fitting function for the prompt photons.
In  section~\ref{sec3}, 
we detail the results for the simulated elliptic flows for both the charged particles and direct photons at RHIC
and LHC. Our final conclusions are summarized in section~\ref{sec:con}.

\section{Electromagnetic radiation in hot QCD matter}
\label{sec2}

\subsection{Hadronic Photon Rates}

Thermal electromagnetic emissions at low and intermediate mass and $q_T$ are involved due to the many 
reaction processes involving hadrons and the strong character of their interactions.  The 
only organizational principles are broken chiral symmetry and gauge invariance, both of
which are difficult to assert  in reaction processes with hadrons in general.   If hadrons
thermalize with the pions and nucleons as the only strongly stable constituents,  then there
is a way to systematically organize the electromagnetic emissivities by expanding them
not in terms of processes but rather in terms of final hadronic states.  The emissivities
are then emmenable to spectral functions by chiral reduction. These spectral functions
are either tractable from other experiments or emmenable to resonance saturation as we now
briefly detail. 

For a hadronic gas in thermal equilibrium the number of photon produced per unit four 
volume and unit three momentum can be related to the electromagnetic current-current 
correlation function \cite{Ste96}

\beqa
\frac{q^0 dN_\gamma}{d^3q}=-\frac{\alpha_{em}}{4\pi^2}\,{\bf W}(q)\,,
\eeqa
with $q^2=0$ and
\beqa
{\bf W}(q)=\int d^4x e^{-iq\cdot x}\Tr\left( e^{-({\bf H}-F)/T} {\bf J}^\mu(x) {\bf J}_\mu(0) \right)\,.
\eeqa
In the above expression ${\bf J}_\mu$ is the hadronic part of the electromagnetic current, 
${\bf H}$ is the hadronic Hamiltonian and $F$ is the free energy.  The trace is over a 
complete set of stable hadronic states for temperatures below $T_c$, e.g. pions and nucleons.
From the spectral representation and symmetry  we can re-express the correlator in terms of 
the absorptive part of the time-ordered correlation function

\beqa
\label{EXACT}
{\bf W}(q)=\frac{2}{1+e^{q^0/T}}\,\Im\,i\int d^4x e^{iq\cdot x}\Tr\left(e^{-({\bf H}-F)/T} T^*{\bf J}^\mu(x) {\bf J}_\mu(0)\right)\,.
\eeqa
 At RHIC and LHC 
the heat bath is net baryon free. The Feynman correlator in (\ref{EXACT}) can be expanded in terms
of final pion states at finite temperature and zero baryon chemical potential

\beqa
{\bf W}^F(q)={\bf W}_{0\pi}+\int d\pi_1 {\bf W}_{1\pi}+ \frac{1}{2!}\int d\pi_1 d\pi_2 {\bf W}_{2\pi}+\cdots
\label{WWW}
\eeqa
with the pion thermal phase space factors

\beqa
d\pi_i = \frac{d^3k_i}{(2\pi)^3}\frac{n(E_i)}{2E_i}\,\,.
\eeqa
We have defined 

\beqa
{\bf W}_{n\pi}&=&i\int d^4x e^{iq\cdot x}
\langle \pi^{a_1}(k_1)...\pi^{a_n}(k_n)\vert T^* {\bf J}^\mu(x) {\bf J}_\mu(0)\vert \pi^{a_1}(k_1)...\pi^{a_n}(k_n)\rangle 
\eeqa
with the sum over isospin subsumed.  
The first contribution in (\ref{WWW}) vanishes for real photons since the heat bath is
stable against spontaneous photon emission. 
The next two terms, ${\bf W}_{1\pi}$ and ${\bf W}_{2\pi}$, can be reduced to measurable vacuum correlators~\cite{Ste96}, e.g.

\beqa
{\bf W}^F_{1\pi}(q,k)&=&\frac{12}{f_\pi^2}q^2\text{Im} {\bf \Pi}_V(q^2)\nn
&-&\frac{6}{f_\pi^2}(k+q)^2\text{Im} {\bf \Pi}_A \left( (k+q)^2\right) + (q\to -q)\nn
&+&\frac{8}{f_\pi^2}\left( (k\cdot q)^2-m_\pi^2 q^2\right) \text{Im} {\bf \Pi}_V(q^2)\times\text{Re} \Delta_R(k+q)+(q\to-q)\nn
\label{eq:lin_in_meson1}
\eeqa
where $\text{Re}\Delta_R$ is the real part of the retarded pion propagator, 
and ${\bf \Pi}_V$ and ${\bf \Pi}_A$ are the transverse parts of the VV and AA 
correlators.  Their spectral functions are related to both $e^+e^-$ annihilation 
and $\tau$-decay data as was compiled in~\cite{Hua95}.  
The two-pion reduced contribution ${\bf W}_{2\pi}$ is more involved.
Its full unwinding can be found in~\cite{Ste96,Ste97,Lee98,Dus07,Dus09,Dus10,Lee14}.

\begin{figure}
    \begin{center}
        \includegraphics[width=0.6\textwidth]{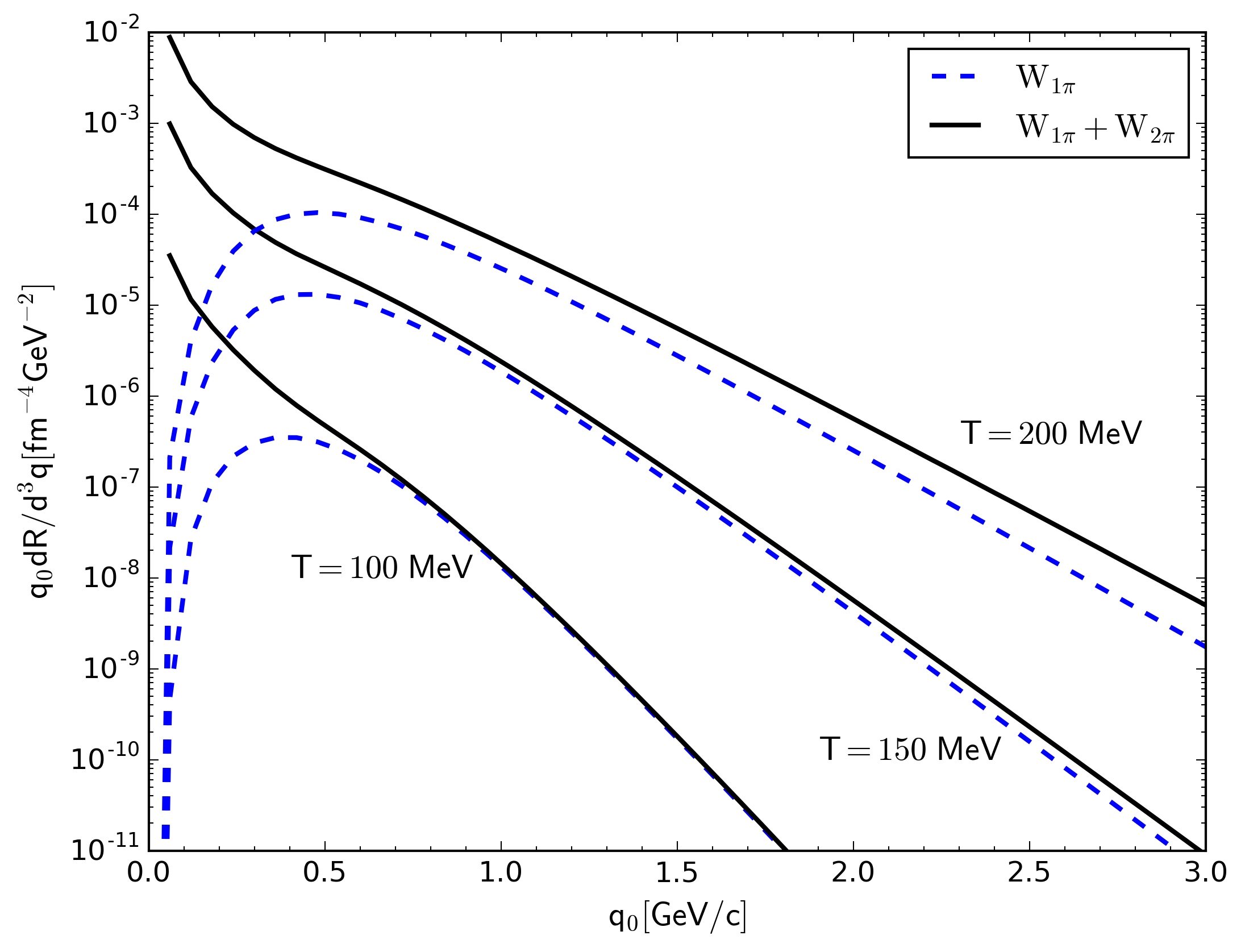}
       \end{center}
         \caption{Photon emission rates from $W_{1\pi}$ and $W_{2\pi}$ with $\mu_\pi=0$ for $T=$100 MeV, 150 MeV and 200 MeV.}
    \label{fig_rate}
\end{figure}

\begin{figure}
    \begin{center}
        \includegraphics[width=0.6\textwidth]{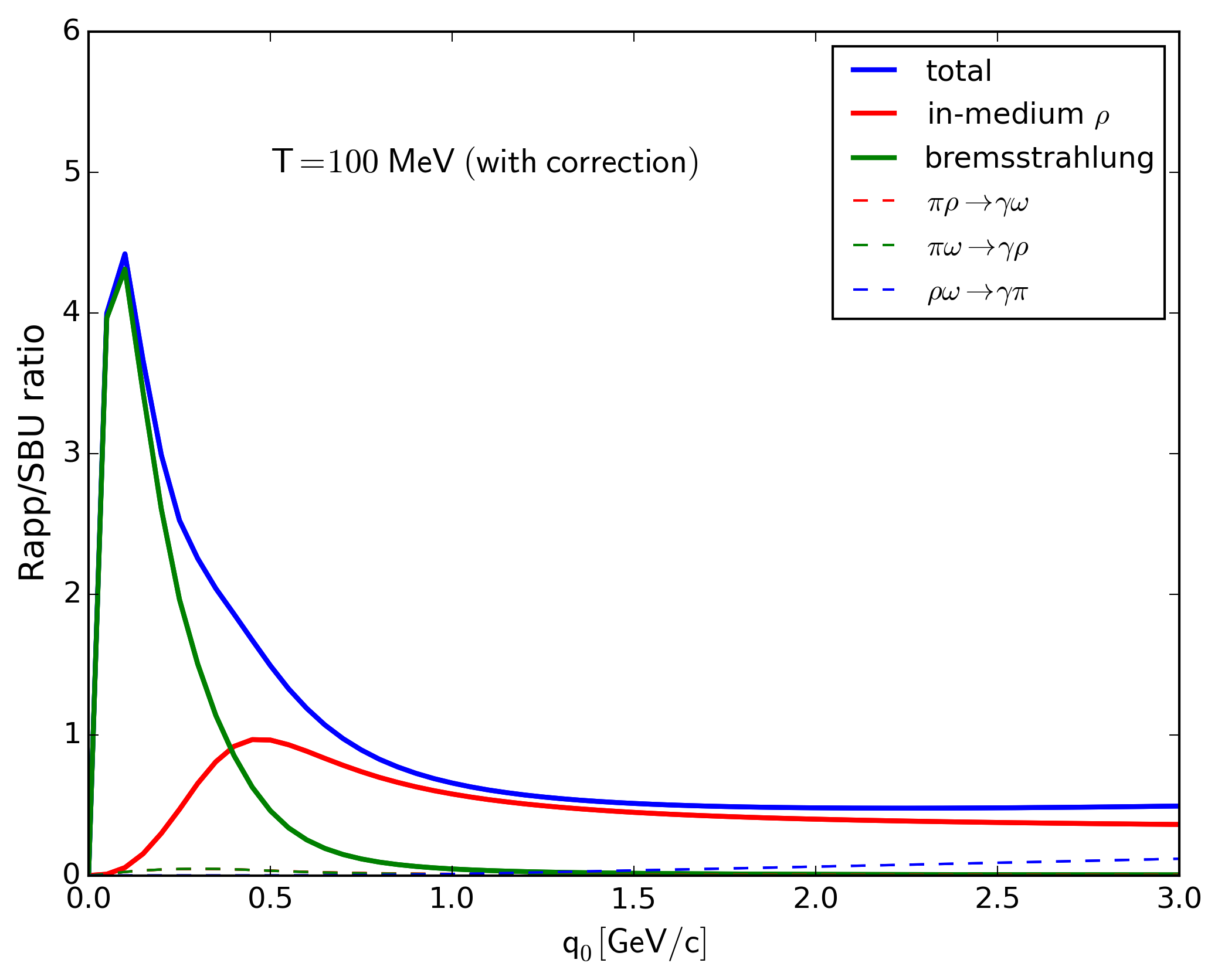}
       \end{center}
         \caption{Ratio of the thermal photon rates used in~\cite{RAPP} (Rapp) to our corrected rates (SBU)  for T=100 MeV. Note that the prompt photon contribution is not included in the thermal photon rate.}    
    \label{fig_comp}
\end{figure}

In Fig.~\ref{fig_rate} we show the  photon rates from the hadonic gas upto the  two-pion contribution without pion chemical potential
and with zero baryon chemical potential. The two pion contribution $W_{2\pi}$ which includes $\pi\pi\rightarrow \rho\gamma$ and $\rho\rightarrow\pi\pi\gamma$ processes dominates at low $q_0$ as discussed in Dusling and Zahed~\cite{Dus10}. Note that our two-pion results are reduced compared to their results due to a  (corrected) reduced phase space in their numerical analysis. 
In Fig.~\ref{fig_comp} we compare the  different rate contributions used by Rapp and collaborators~\cite{RAPPPI} to our corrected rates at $T=100$ MeV.  Our corrected 2-pion contribution which includes the Bremsstrahlung is about 3 times smaller at the
highest point near treshold. We note that in the diagrammatic analysis in~\cite{RAPP} of the photon Bremsstrahlung,  current
conservation of the photon polarization function is enforced by hand, while in our analysis it  is  manifestly satisfied 
by  the chiral reduction scheme.

\begin{figure}
    \begin{center}
        \includegraphics[width=0.6\textwidth]{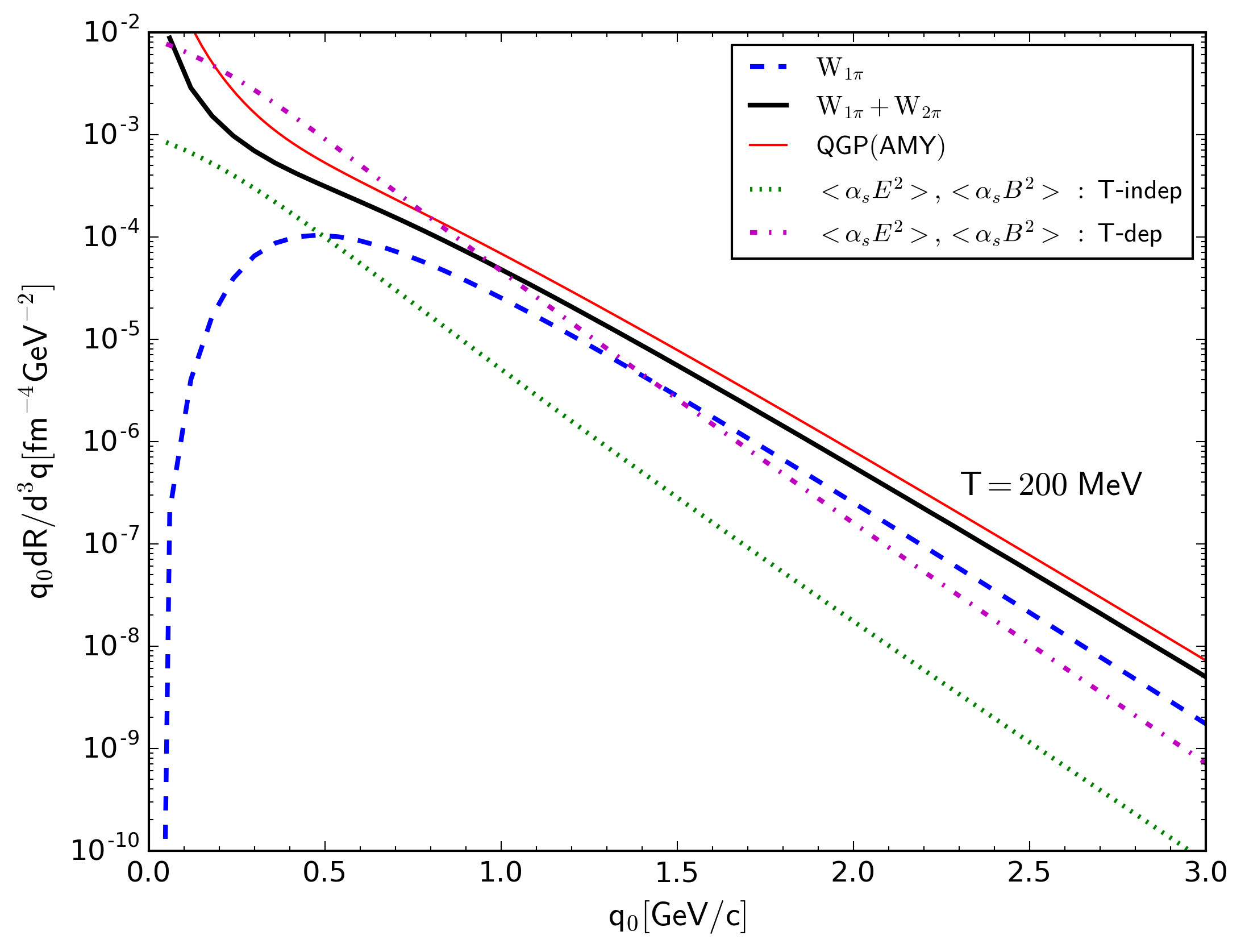}
       \end{center}
         \caption{Photon emission rates from $W_{1\pi}$ and $W_{2\pi}$ with $\mu_\pi=0$ for $T=$ 200 MeV are compared
         to the AMY rates  (AMY; Arnold et al.~\cite{AMY} with $N_f=3$)  for $T=$200 MeV. Also shown are the non-perturbative 
         soft gluon corrections from~\cite{Lee14,SOFT}.}    
    \label{fig_rate_c}
\end{figure}

\begin{figure}
    \begin{center}
        \includegraphics[width=0.6\textwidth]{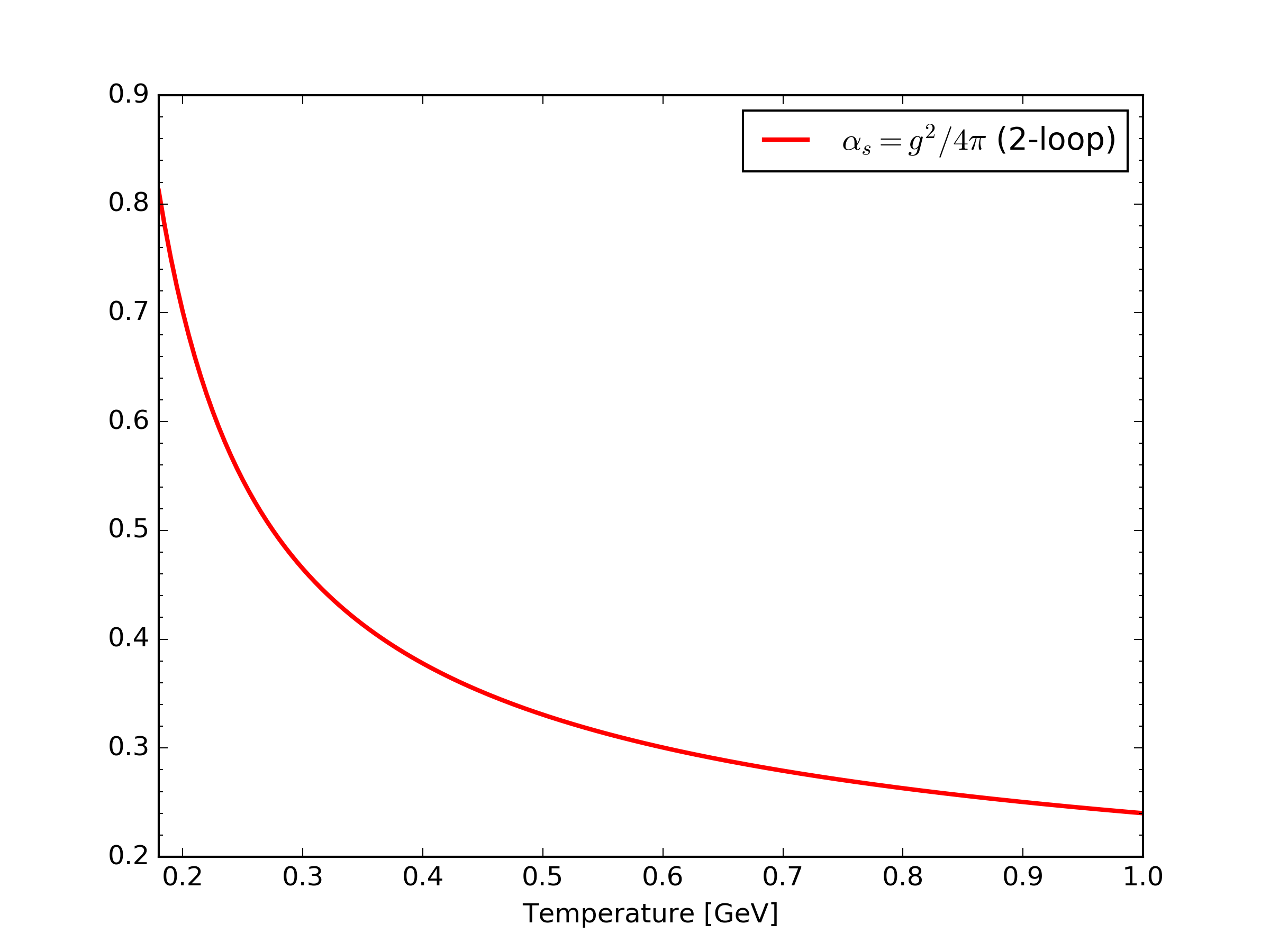}
       \end{center}
         \caption{Running strong coupling constant $\alpha_s$ up to 2-loop order used in the AMY rates~\cite{Chet98}.}    
    \label{fig_alpha_s}
\end{figure}



\subsection{QGP Emission}

There has been great progress in the calculation of the QGP photon rates in 
QCD both at leading~\cite{AMY} and next-to-leading order~\cite{Ghiglieri:2013gia}.  We will not go over the details
of the recent analyses but rather highlights the key points. 
Firstly, the one loop diagram corresponding to $q\overline{q}\to \gamma$
contributes at order $\alpha_s^0$ for dileptons but vanishes at the photon
point due to energy momentum conservation.  One would then expect that the
leading order in $\alpha_s$ contribution will come from two-loop diagrams
corresponding to the annihilation $q+\overline{q}\to\gamma$ and compton
$g+q(\overline{q}) \to q(\overline{q}) + \gamma$ processes.  However these
rates are plagued with collinear singularities~\cite{AMY}. Instead, a complete
leading order photon emission requires the inclusion of collinear bremsstralung
and inelastic pair annihilations  and their subsequent suppression through the
LPM effect. We will refer to NLO QCD calculation as the resummed  QGP (AMY) rates.

In Fig.~\ref{fig_rate_c}, the resummed AMY QGP rates at high temperature~\cite{AMY} are compared to the hadronic rates at
$T=200$ MeV. In Fig.~\ref{fig_alpha_s}, we summarize the running strong coupling constant $\alpha_s$ as a function of temperature
used in assessing the AMY rates. The comparison shows that the AMY rates are substantially higher due to the running up of the coupling
constant, an indication of non-perturbative physics as suggested in~\cite{Lee14,SOFT}, and also
 in~\cite{SEMI}. Non-perturbative
contributions through soft gluon insertions from~\cite{Lee14,SOFT} using the OPE expansion in leading order,
are also shown in Fig.~\ref{fig_rate_c} for comparison.  We note that these contributions are substantial in the Bremsstralung region. In contrast  to the AMY rates, they are finite at zero photon frequency 
to guarantee a finite electric conductivity~\cite{Lee14}. In the hydrodynamical estimates of the emissivities only the hadronic and AMY rates will be retained. It is worth noting that the AMY rates used in the photon emissivities by the McGill group~\cite{Gal15} makes use of a fixed $\alpha_s\approx 0.3$ and are therefore lower than the ones we used with a running $\alpha_s$ as in Fig.~\ref{fig_alpha_s}.

%
%

The lattice EoS \cite{Laine06cp} includes a rapid crossover followed by an
interpolation into the hadronic resonance gas phase.  Even though there is no
true phase transition, we choose $T_{\rm crit}=190$ MeV to allow for a  switch
from partonic to hadronic electromagnetic emission.  This choice of $T_{\rm crit}$ does not
affect the hydrodynamic evolution.

\section{Azimuthal Anisotropy}
\label{secani}

The distribution of the emitted photons follow from the integrated
space-time hydrodynamically evolved emission rates within
the freeze-out volume

\beqa
\frac{d^3 N_\gamma}{q_T dq_T dy d\phi} (q_T, y, \phi)=
 \int_{\tau_0}^{\tau_{f,o}}\!\!\!\! \tau d\tau \int_{-\infty}^{\infty} \!\!\! d\eta \int_0^{r_{\rm max}}\!\!\!\! r dr
  \int_0^{2\pi} \!\!\! d\theta\; 
 \,\left[ q^0 \frac{d R_\gamma}{d^3q} (q=\vec q\cdot \vec u;T,\mu_B,\mu_\pi) \right]\,\Theta(T>T_{\rm FO})
\label{eq:rate}
\eeqa
Here $R_\gamma\equiv dN_\gamma/d^4x$ is the photon production rate, i.e., the
number of direct photons per unit four-volume in the local rest frame of the fire ball.
The hydrodynamical
evolution  is based on a numerical code developed by one of us~\cite{Mazeliauskas:2015vea}. Its key parameters
will be briefly detailed in the next section. The rapidity is $y=\frac 12 {\rm ln}((E+q_L)/(E-q_L))$, the proper time is $\tau=\sqrt{t^2-z^2}$ and the spatial rapidity is $\eta=\frac 12 {\rm ln}((t+z)(t-z))$~\cite{Bjo83}.

The elliptic flow and higher harmonics $v_n (q_T, y)$ in each event follow by expanding (\ref{eq:rate}) in  Fourier components. In general
the Fourier series requires sines and cosines, or amplitudes and phases
\beqa
\frac{d^3 N_\gamma}{q_T dq_T dy d\phi}
=\frac{1}{2\pi } \frac{d^2 N_\gamma}{q_T d q_T dy} \left( 1+  \sum_{n=1}^\infty  v_{n \gamma} (q_T,y)  e^{in (\phi-\Psi_{n\gamma}(q_T,y) ) } \right)  + {\rm c.c.} \, ,
\eeqa
where ${\rm c.c.}$ denotes complex conjugation. 
The amplitude $v_{n\gamma}(q_T,y)$ is real and positive semi-definite, and the phase is real.
The amplitudes and  phases of the photon yield are combined into a single
complex  event-by-event flow coefficient 
\st
\label{complexflow}
\mathcal V_{n\gamma}(q_T,y) \equiv v_{n\gamma}(q_T,y) e^{-in \Psi_{n\gamma}(q_T,y) } \, .
\stp
The integrated event-by-event pion yield is also expanded in 
a Fourier series~\footnote{The event-by-event pion yield
    a contribution of direct pions and feed-down pions from resonance decays.
We will work with the direct pion yield in what follows. }
\beqa
\frac{dN_\pi}{d\phi}
=\frac{1}{2\pi } N_{\pi} \left( 1+  \sum_{n=1}^\infty  v_{n \pi} e^{in (\phi-\Psi_{n\pi} ) } \right)  + {\rm c.c.} \, ,
\eeqa
and the complex flow coefficients, $\mathcal V_{n\pi}=v_{n\pi}e^{-in\Psi_{n\pi}}$ are defined in analogy with \Eq{complexflow}.
The measured ``photon elliptic flow" is the event-averaged correlation between 
the photon yield and integrated charged hadron yields which define
the $n$-th order reaction plane. We will use the thermal pion yield as 
a proxy for this event plane, and therefore the photon elliptic flow 
is defined in the simulation as
\st
v_{2\gamma}\{2\}(q_T, y)  \equiv 
\frac{\llangle \V_{n\gamma}(q_{T},y) \V_{n\pi}^* \rrangle }
{ \sqrt{ \llangle |\V_{n\pi}|^2 \rrangle } }  \, .
\stp
where the bracket refers to event averaging.
We will use this operational definition of the photon elliptic flow in what follows.

\section{Brief of Hydrodynamics}
\label{sechydro}

The collision region is modeled using a relativistic hydrodynamic simulation
tuned in order to reproduce hadronic observables.  In this section we briefly
discuss the model, including the initial conditions and equation of state
(EoS),  but leave the technical details to the literature~\cite{Mazeliauskas:2015vea}.  
We use the Phobos Glauber Monte Carlo \cite{Alver:2008aq}
model to initialize the entropy density $s(\tau_o,\x)$ in the transverse plane
at an initial proper time $\tau_o$ according to a two component model. 
Briefly, for the $i_{\rm th}$  participant we assign a weight

\st
A_i \equiv \kappa \left[ \frac{ (1- \alpha) }{2}   + \frac{\alpha}{2} (n_{\rm coll})_i \right] \, ,
\stp
where $\alpha=0.11$ is adjusted to reproduce the mean multiplicity 
versus centrality.
At the LHC we take $\kappa = 27.0$, while at RHIC we take $\kappa  = 15.5 $ so that
$\kappa_{\rm LHC}/\kappa_{\rm RHIC} =  1.74 $ which matches the ratio of multiplicities at the two colliding systems. 
$(n_{\rm coll})_i$ is the number of binary collisions experienced by the $i_{\rm th}$
participant. The entropy density in the transverse plane at initial time $\tau_o$ and transverse
position $\x=(x,y)$ is taken  to be
\st
s(\tau_o,\x)  = \sum_{i {\rm parts}} s_{i}(\tau_o, \x - \x_i) \, ,
\stp
where $\x_i = (x,y)$ labels the transverse coordinates of the $i_{\rm th}$ participant, and
\st
s_{i}(\tau_o, \x) = A_i \, \frac{1}{\tau_o (2\pi \sigma^2) } e^{- \frac{x^2}{2\sigma^2} -\frac{y^2}{2\sigma^2} } \, ,
\stp
with $\sqrt{2}\sigma = 0.7\,{\rm fm}$.
The parameters $\kappa$ and $\alpha$ are comparable to those used in~\cite{Qiu:2013wca}.

Table~\ref{tab1} shows the choice
of parameters~\cite{Adl04,Adl03} for PHENIX (\textsuperscript{197}Au+Au at $\sqrt{s_{NN}}=200$ GeV). Again, $N_{\rm part}$ is the number of participants (nucleons) and $N_{\rm coll}$ is the number of  collisions among nucleons.   The nucleon-nucleon inelastic cross section is $\sigma_{\rm inel}^{NN}=40$ mb \cite{PHE03a}, and the entropy per wounded nucleon is $S_{\rm WN}=15.5$.
Table~\ref{tab2} shows the choice of parameters~\cite{ALICE13,CMS13,ALICE} for ALICE/CMS
(\textsuperscript{208}Pb+Pb at $\sqrt{s_{NN}}=2.76$ TeV). 
The inelastic nucleon-nucleon cross section is $\sigma_{\rm inel}^{NN}=64$ mb~\cite{CMS14}, and the entropy per wounded nucleon is now $S_{\rm WN}=27$. In order to take into account the event-by-event fluctuations, we performed 300 runs for each centrality region. 
For PHENIX data the freeze-out temperature is set to $T_{\rm FO}=137$ MeV, while 
for ALICE data the freeze-out temperature is set slightly lower with $T_{\rm FO}=131$ MeV.

\begin{figure}[h]
\centering
\includegraphics[scale=0.4]{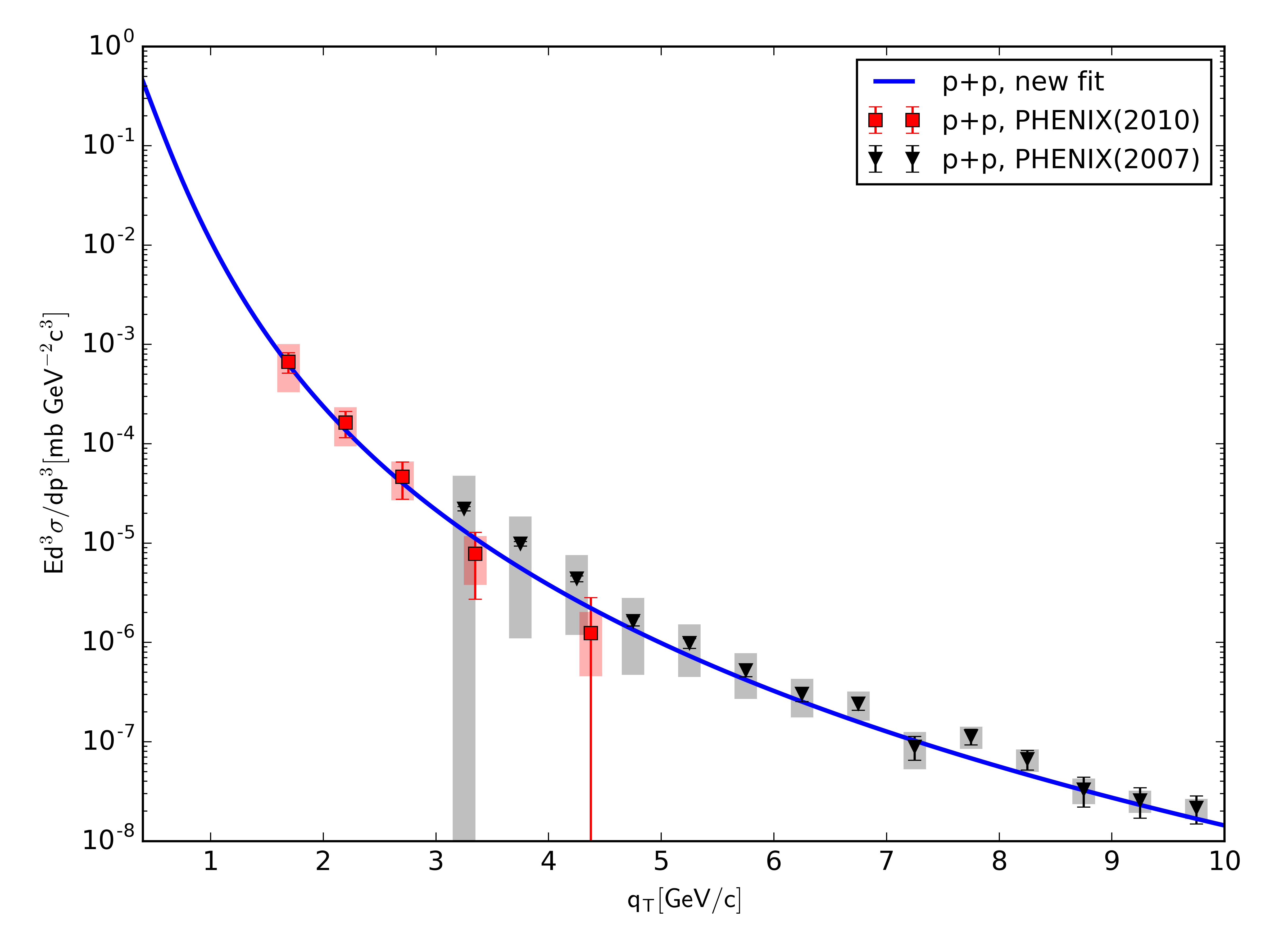}
\caption{Prompt photon spectra fitted to PHENIX p$+$p \cite{PHE07,PHE10}}
\label{prompt_pp_RHIC}
\end{figure}

\begin{figure}[h]
\centering
\subfigure[~0-20\%]{\includegraphics[scale=0.4]{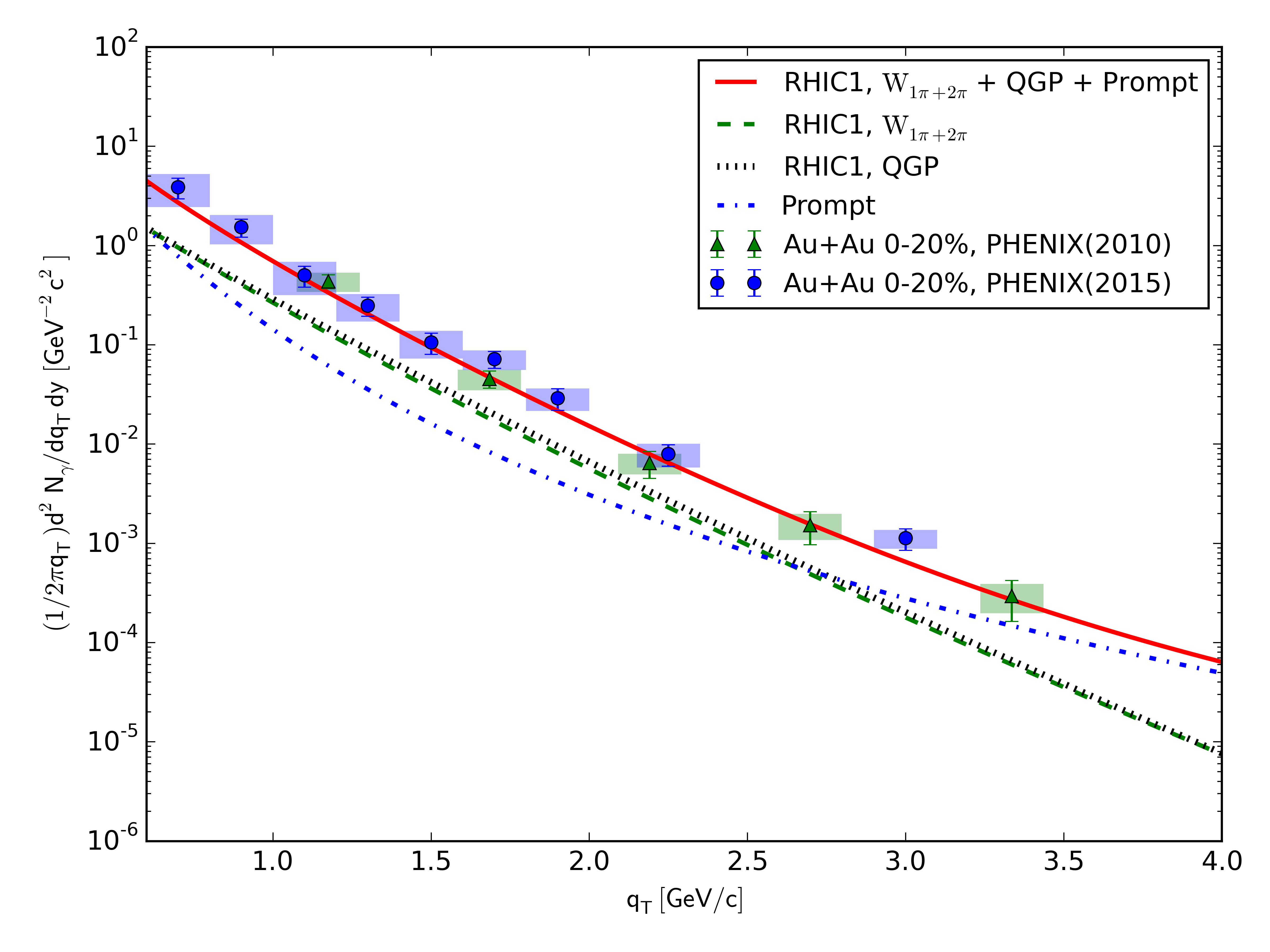}}
\subfigure[~20-40\%]{\includegraphics[scale=0.4]{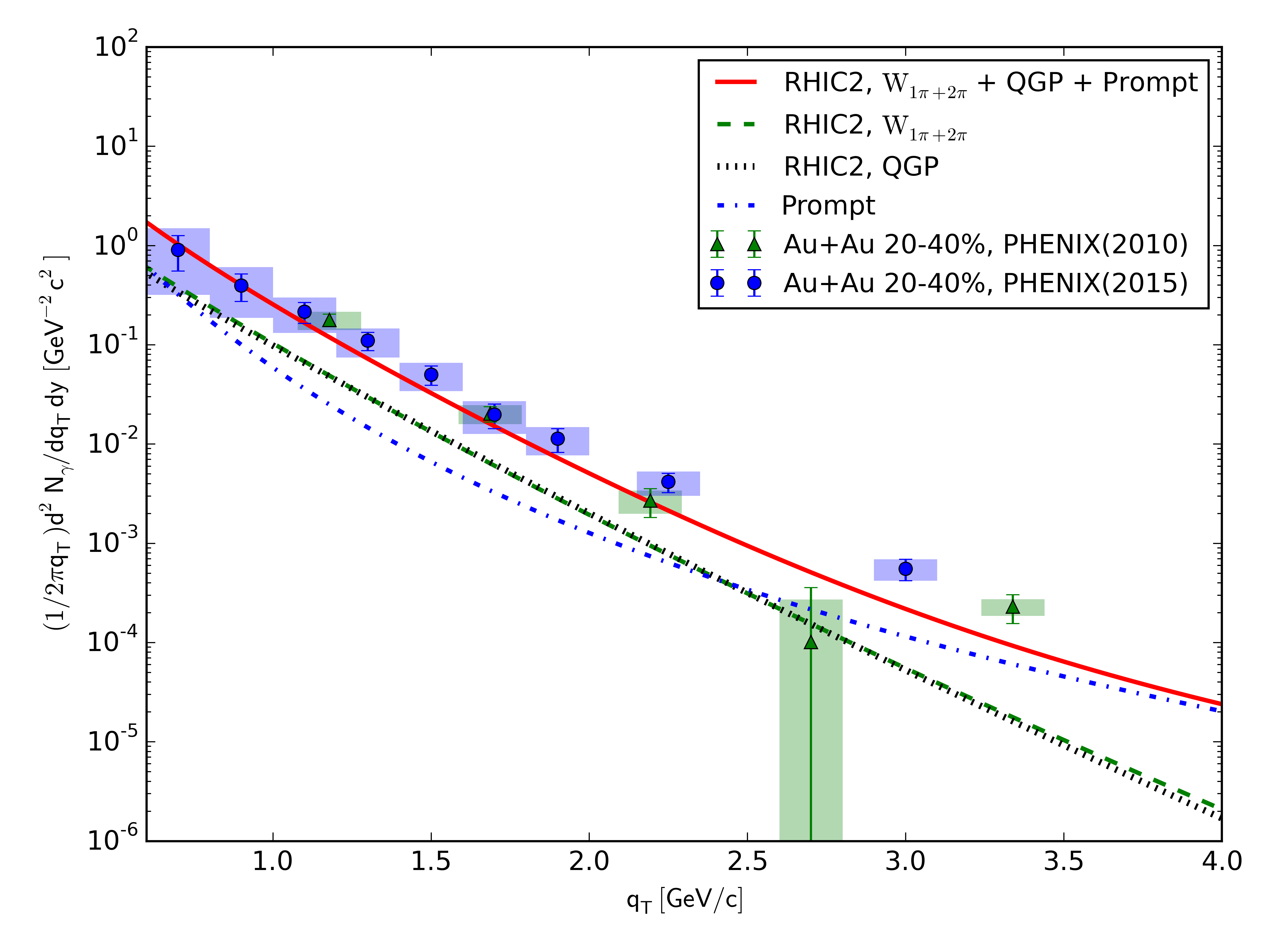}}
\subfigure[~40-60\%]{\includegraphics[scale=0.4]{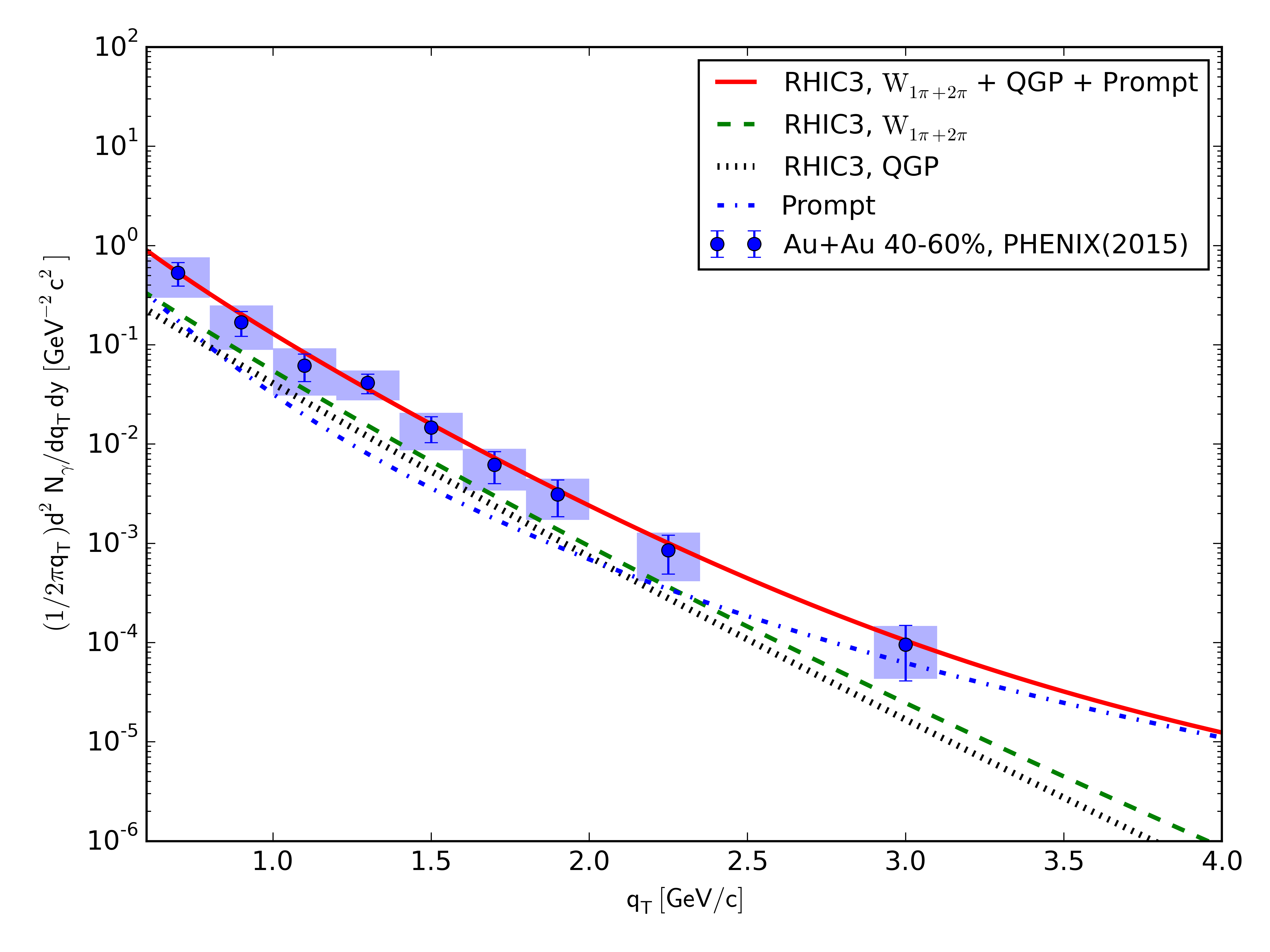}}
\caption{Spectra of Direct photon for RHIC. Experimental data for PHENIX Au$+$Au is taken from~\cite{PHE10, PHE15}. See text.}
\label{spectra_rhic}
\end{figure}

\begin{figure}[h]
\centering
\subfigure[~0-20\%]{\includegraphics[scale=0.4]{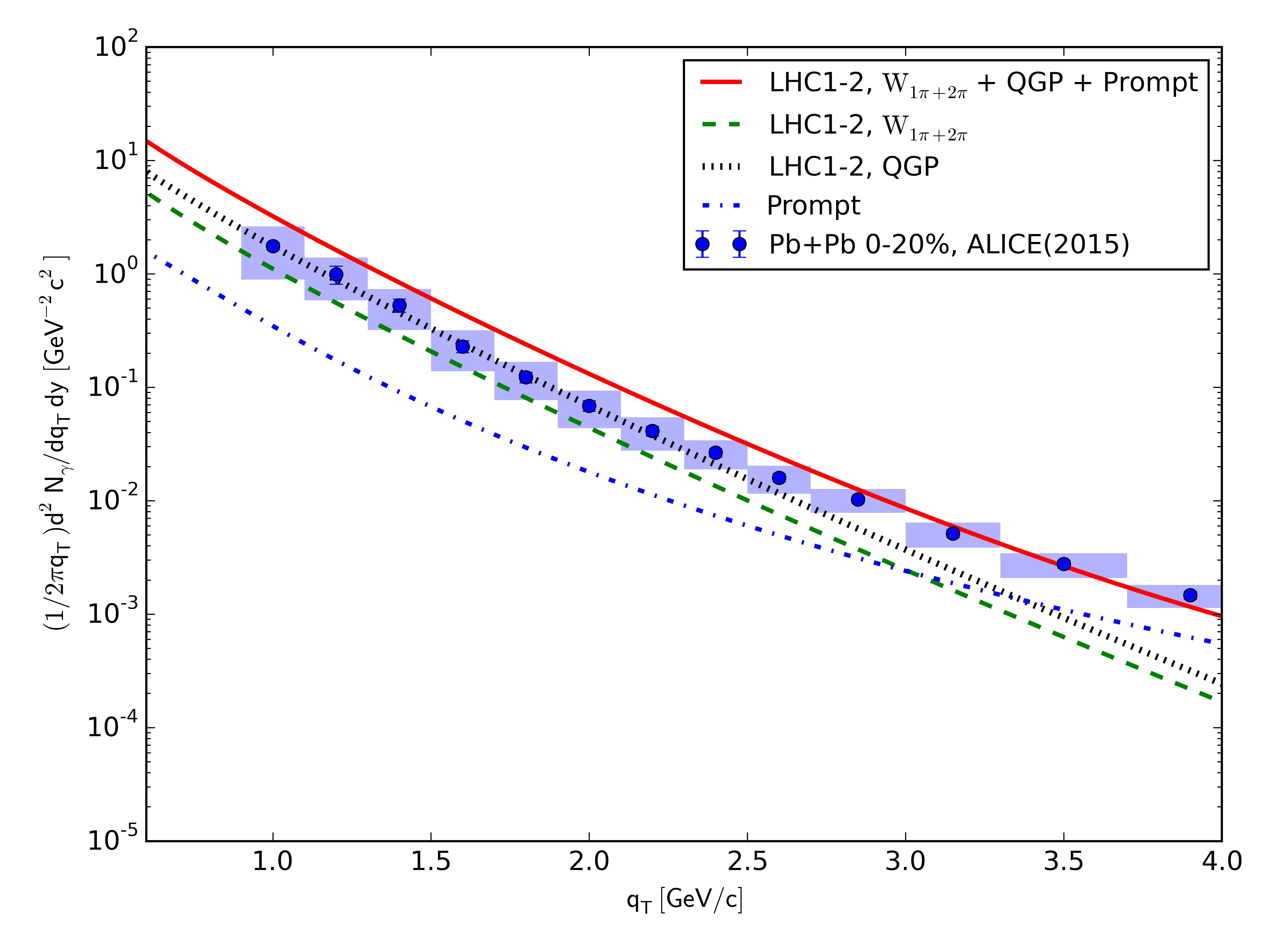}}
\subfigure[~20-40\%]{\includegraphics[scale=0.4]{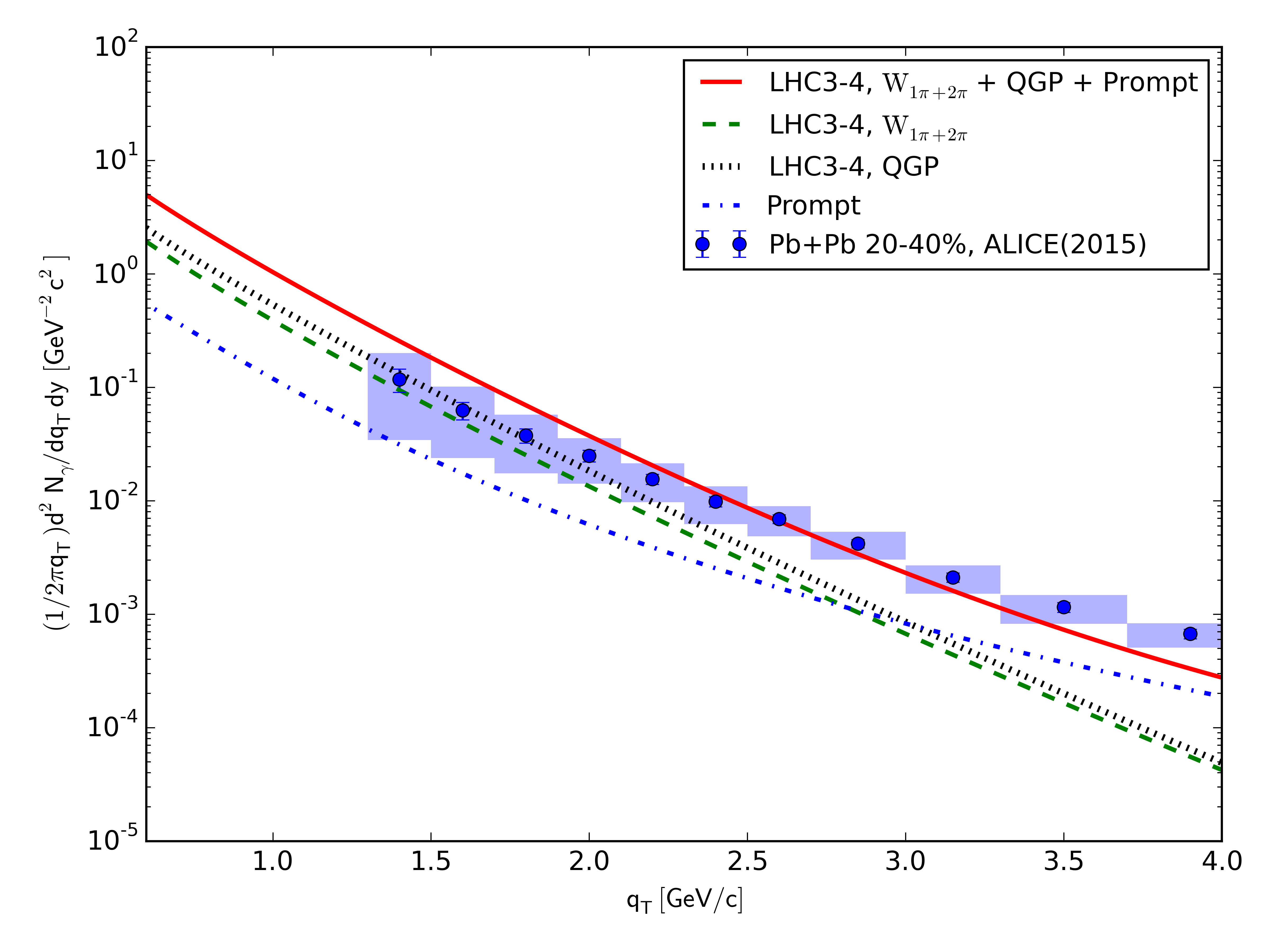}}
\caption{Spectra of Direct photon for ALICE. Experimental data for ALICE Pb$+$Pb is taken from~\cite{ALICE16}. Note that LHC1-2 (LHC3-4) was averaged over 2 centrality ranges, 0-10\% and 10-20\% (20-30\% and 30-40\%) weighted by the number of photons at each centrality bin~\cite{Lohner13}. See text.}
\label{spectra_alice}
\end{figure}

\section{Photon spectra}
\label{sec_pp}

The direct photons include thermal photons as well as prompt photons which are produced by hard parton-parton collisions and can be assumed azimuthally symmetric.
The prompt photon spectrum can be estimated by an empirical model to fit the measured photon spectrum in $p+p$ collisions scaled by the number of binary collisions. 
The empirical model is described in \cite{Sri01,GALE} (and references therein) as
\beqa
q_0\frac{d^3 N^{\rm prompt}_{\gamma}}{d^3 q} = q_0\frac{d^3 \sigma^{\rm pp}}{d^3 q} \frac{N_{\rm coll}}{\sigma^{\rm inel}_{\rm NN}},\label{ap-eq:4}
\eeqa
where $N_{\rm coll}$ is the number of nucleus-nucleus (N-N) collisions and $\sigma^{\rm inel}_{\rm NN}$ is the inelastic scattering cross section in N-N collisions. To parametrize the prompt photon at RHIC for $p+p$ collisions,  we will use the parametrizationcan~\cite{PHE10},
\beqa
q_0\frac{d^3 \sigma^{\rm pp}}{d^3 q} = A \left(1+\frac{q_T^2 }B\right)^{-n} {\rm \frac{mb}{GeV^2 c^{-3}}}.\label{ap-eq:5}
\eeqa
%
%
The parameters $A=2.6955$, $B=0.19943$ and $n=3.0631$ follow by fitting the p-p spectrum in the PHENIX 
experiment~\cite{PHE07, PHE10} as shown in~Fig.~\ref{prompt_pp_RHIC}.

In Fig.~\ref{spectra_rhic} we show our  direct photon spectrum in comparison to the 
PHENIX (\textsuperscript{197}Au+$^{197}$Au at $\sqrt{s_{NN}}=200$ GeV)~\cite{PHE10, PHE15} measurements.
The blue dot-dashed lines correspond to the prompt photon spectrum which is obtained from the p-p spectra in Fig.~\ref{prompt_pp_RHIC} scaled by the number of binary collisions. 
The dashed and dotted lines correspond to our predictions of the thermal photon production for hadronic and QGP contributions.
The red solid line  is the total spectrum of the direct photons which is obtained by summing the thermal and prompt photons.

The prompt $p+p$ spectrum has not been reported by ALICE~\cite{ALICE16}. However, a theoretical analysis of 
the PHENIX data using perturbative QCD at NLO shows a strong sensitivity to $\sqrt{s}$ in extrapolating the prompt spectra
from RHIC to LHC~\cite{AURENCHE,PAQUET}. The $p_T$ dependence of the prompt spectrum at $\sqrt{s}= 2.76\,{\rm TeV}$
in~\cite{AURENCHE,PAQUET} can be reproduced by the parametrization (\ref{ap-eq:5}) with the parameter set
$A=0.55269$, $B=0.48304$, and $n=2.6788$. It is substantially higher than the one reported by PHENIX in the same $p_T$ range.
In Fig. \ref{spectra_alice} we show our  direct photon spectrum in comparison to 
ALICE (\textsuperscript{208}Pb+$^{208}$Pb at $\sqrt{s_{NN}}=2.76$ TeV) measurements~\cite{ALICE16}. The description of the curves 
is identical to the one presented for PHENIX above. We note that the prompt yield following from  (\ref{ap-eq:5}) becomes comparable
to our hadronic yield  at small $p_T$. Overall,  our direct
photon yields are  larger than those reported recently by the McGill group for the same $p_T$ ranges
and centralities~\cite{PAQUETX}. This is most likely due to our use of the running coupling constant in the AMY rates which
is larger.


\begin{table}
\caption{(PHENIX) Number of particles and simulation parameters  for $v_2$ calculation in Figs.~\ref{fig1} and \ref{fig2}. The experimental data are taken from PHENIX~\cite{Adl04,Adl03}. 
 $N_{\rm part}$ is the number of participants (nucleons) and $N_{\rm coll}$ is the number of  collisions among nucleons. The centrality is mainly determined by $N_{\rm part}$. The number of produced charged particles ($N_{\pi^\pm}$ and $N_{p+\bar p}$) in our simulations are much smaller than those in the experiments because our simulations were done only upto the freeze-out temperature.
 }
\label{tab1}
\begin{center}
\begin{tabular}{|c|c|c|c|c||c|c|c|c|}
\hline
\multicolumn{5}{|c||}{simulation} & \multicolumn{4}{c|}{experiment} \\
\cline{1-9}
run ID &   $b$(fm) & $N_{\rm part}$ & $N_{\rm coll}$ & $N_{\pi^\pm}$ (direct) &   $N_{\rm part}$ & $N_{\rm coll}$ 
& $N_{\pi^\pm}$ (all)  & centrality \\
\hline
RHIC1 &   6.08$\pm$0.01 & 214.67$\pm$14.71 &  517.66$\pm$59.73 & 149.73$\pm$11.54   & 215.3$\pm$5.3 &   532.7$\pm$52.1 & 341.2$\pm$30.0   & 15-20\% \\
RHIC2 &   8.65$\pm$0.01 & 114.43$\pm$13.86 & 213.62$\pm$40.17 & 73.00$\pm$9.76 &      114.2$\pm$4.4   & 219.8$\pm$22.6 & 171.4$\pm$16.6 &   30-40\% \\
RHIC3 &   9.85$\pm$0.01 & 74.52$\pm$13.19 & 115.75$\pm$30.30 & 44.92$\pm$8.67 &           74.4$\pm$3.8  & 120.3$\pm$13.7 & 107.8$\pm$10.8 &   40-50\% \\
\hline
\end{tabular}\\
\end{center}
\end{table}

\begin{table} 
\caption{(ALICE/CMS) Number of particles and parameters for the simulations including higher centrality as summarised in Figs~\ref{fig3} and \ref{fig4}. The experimental data are taken from the ALICE and CMS experiments \cite{ALICE13,CMS13,ALICE}.
$N_{\rm part}$ is the number of participants (nucleons) and $N_{\rm coll}$ is the number of  collisions among nucleons. 
The centrality is mainly determined by $N_{\rm part}$. The produced charged particles ($N_{\pi^\pm}$ and $N_{p+\bar p}$) in our simulations are  smaller than those in the experiments because our simulations were done only upto the freeze-out temperature.}
\label{tab2}
\begin{center}
\begin{tabular}{|c|c|c|c|c||c|c|c|c|}
\hline
 \multicolumn{5}{|c||}{simulation} & \multicolumn{4}{c|}{experiment} \\
\cline{1-9}
 run ID &   $b$(fm) & $N_{\rm part}$ & $N_{\rm coll}$ & $N_{\pi^\pm}$  (direct) & $N_{\rm part}$ & $N_{\rm coll}$ 
 & $N_{\pi^\pm} $  (all) & centrality \\
\hline
LHC1 &   4.27$\pm$0.01 & 330.00$\pm$12.24 & 1315.33$\pm$104.34 & 549.27$\pm$26.62  &   329$\pm$3 & $-$ & 1210$\pm$84 &   5-10\% \\
LHC2 &   6.49$\pm$0.01 & 239.58$\pm$16.12 &   811.27$\pm$96.10   & 369.67$\pm$29.62  &   240$\pm$3 & $-$ & $-$   & 15-20\% \\
LHC3 &   8.13$\pm$0.01 & 170.28$\pm$16.16 &   493.37$\pm$79.87   & 246..09$\pm$27.74 &   171$\pm$3 & $-$ & $-$   & 25-30\% \\
LHC4 &   9.49$\pm$0.01 & 117.34$\pm$15.79 &   278.78$\pm$53.28   & 157.31$\pm$22.75  &    118$\pm$3 & $-$ & $- $   & 35-40\% \\
\hline
\end{tabular}\\
\end{center}
\end{table}

\section{Numerical Results : Elliptic Flow}
\label{sec3}

\subsection{Elliptic Flow at RHIC}

\begin{figure}[h]
\centering

\subfigure[ ~0-20\% centrality]{\includegraphics[scale=0.4]{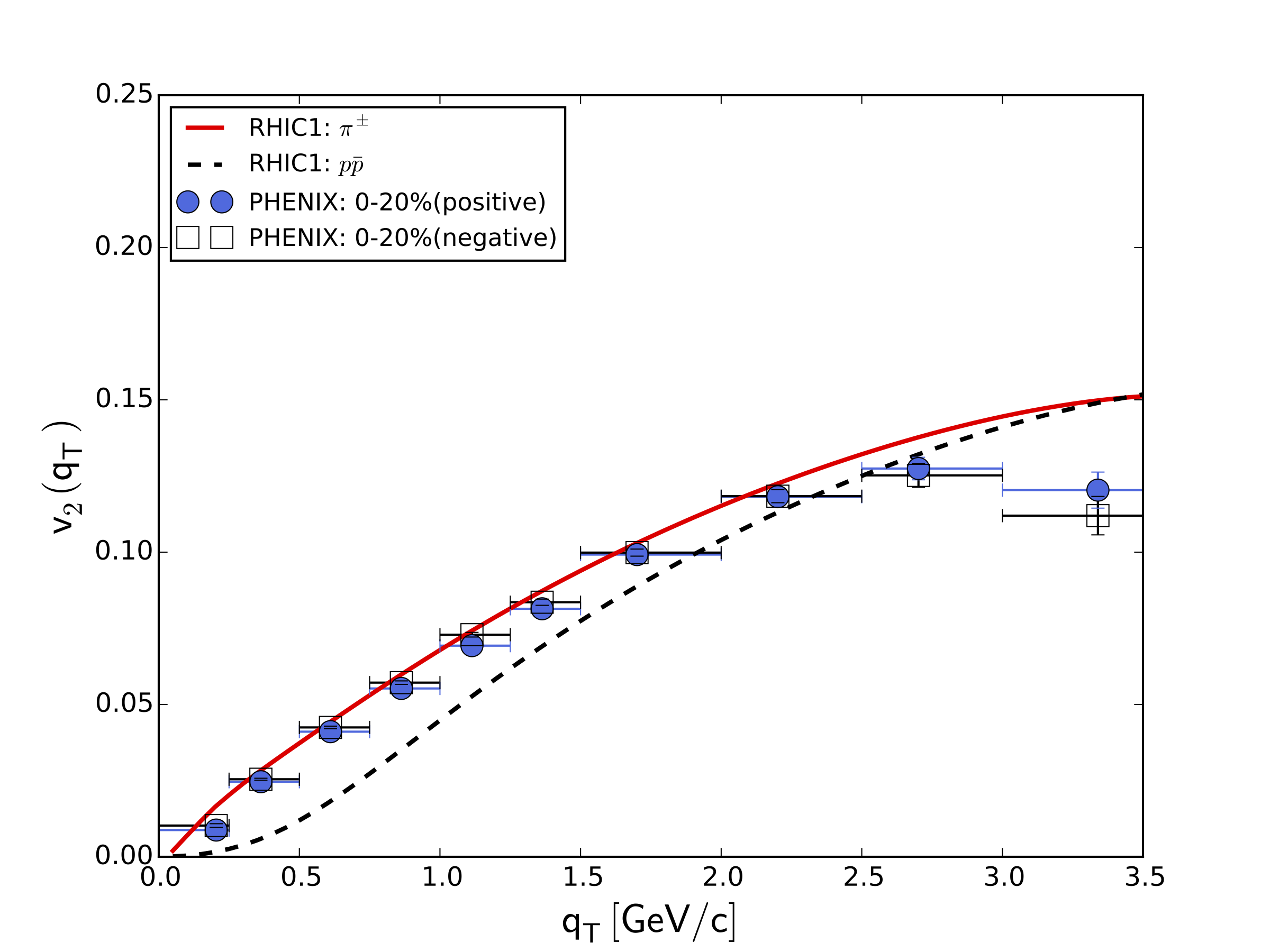}}
\subfigure[ ~20-40\% centrality]{\includegraphics[scale=0.4]{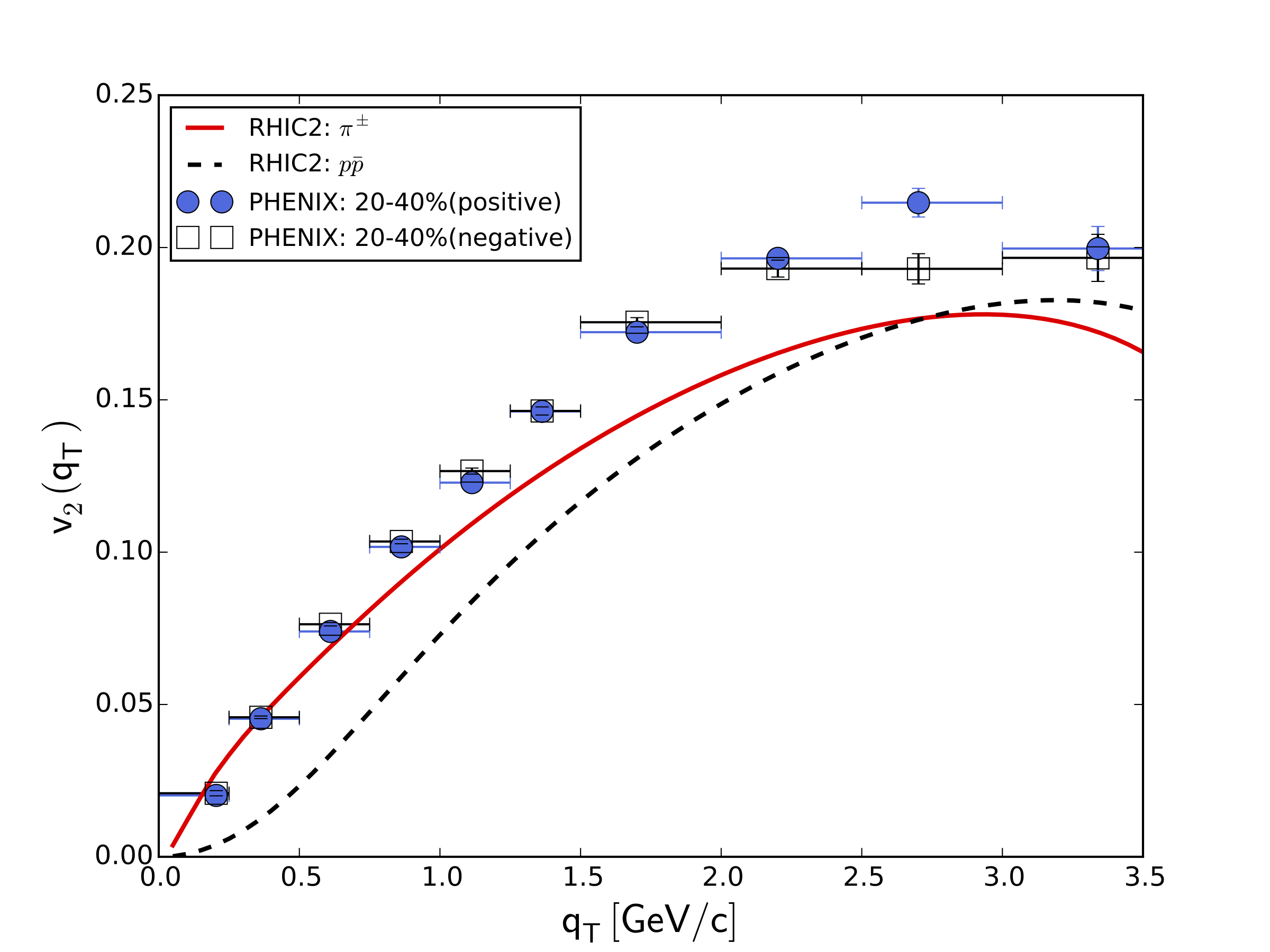}}
\subfigure[ ~40-60\% centrality]{\includegraphics[scale=0.4]{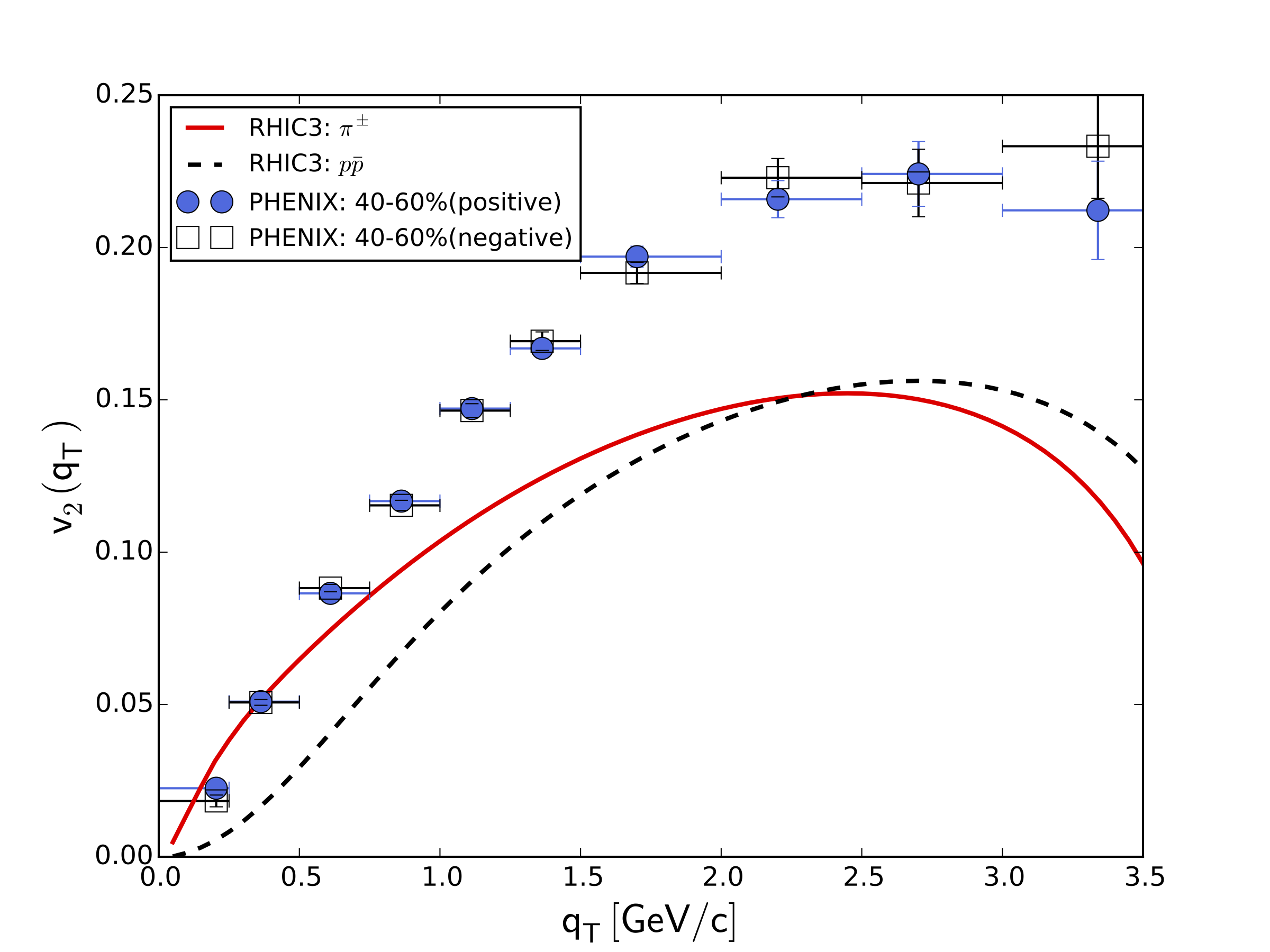}}

\caption{(PHENIX: charged particles) Elliptic flow $v_2$ of charged particles ($\pi^\pm, p, \bar p$) for PHENIX~\cite{Adl03}.  The parameters for the simulations are summarised in Table~\ref{tab1}. In this figure, we display the best-fit results in a given centrality region.
}
\label{fig1}
\end{figure}

In Fig.~\ref{fig1}, the simulation results for PHENIX $v_2$ of charged particles are summarized and compared with  the experimental results~\cite{PHE03a,Adl03,Adl04}.
The parameters used in our simulations are sumarised in Table~\ref{tab1}. The nucleon-nucleon inelastic cross section $\sigma_{inel}^{NN} = 40.0$ mb is consistent  with the referred cross section for the Glauber model,  $\sigma_{inel}^{NN} = 42\pm 3$~\cite{PHE03a}. 
Note that the numbers of produced charged particles ($N_\pi$ and $N_p$) in our simulations are smaller than those in experiments because our simulations were done only upto the freeze-out temperature.  Resonance decays are not included. 
For PHENIX, our simulated elliptic flows of charged particles are somewhat below the experimental data in most of the ranges of the transverse momentum, especially at high transverse momentum.
For low centrality, 0-20\% in Fig.~\ref{fig1}~(a), our best-fit simulation results are consistent with the experimental results. For middle and high centrality in Figs.~\ref{fig1}~(b) and (c), the simulation results are below the experimental data.

\begin{figure}[b]
\centering

\subfigure[~0-20\% centrality]{\includegraphics[scale=0.6]{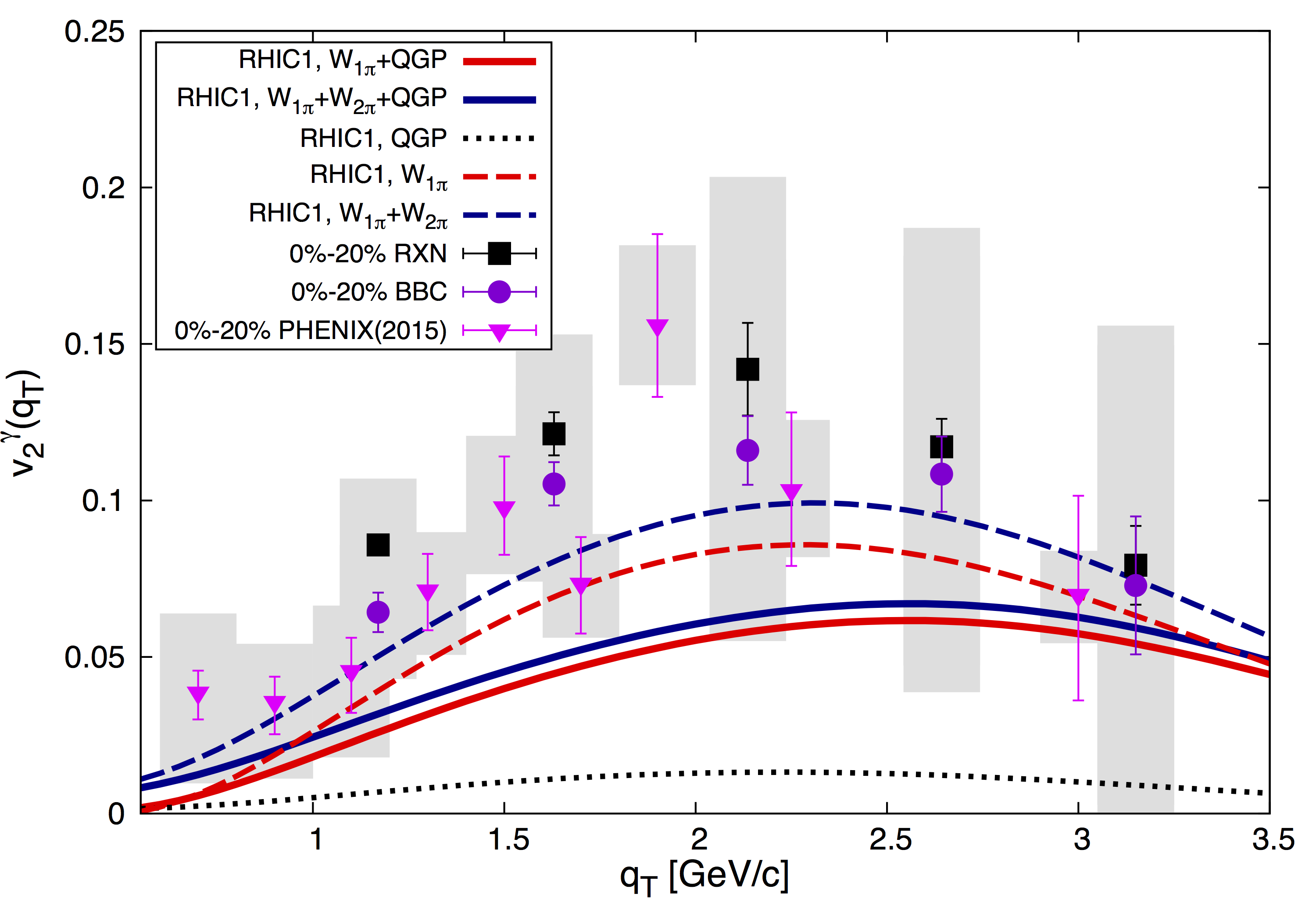}}
\subfigure[~20-40\% centrality]{\includegraphics[scale=0.6]{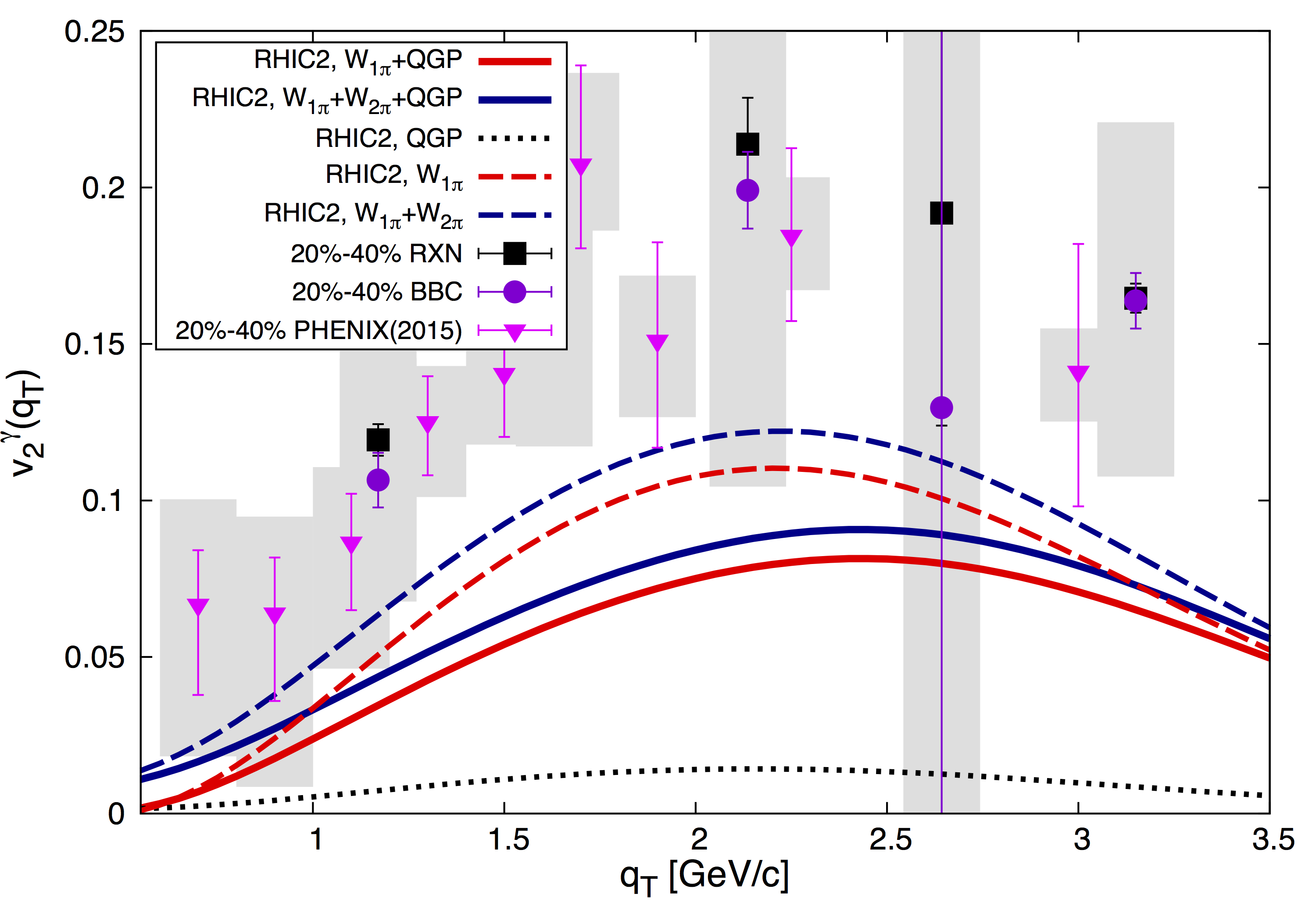}}
\subfigure[~40-60\% centrality]{\includegraphics[scale=0.6]{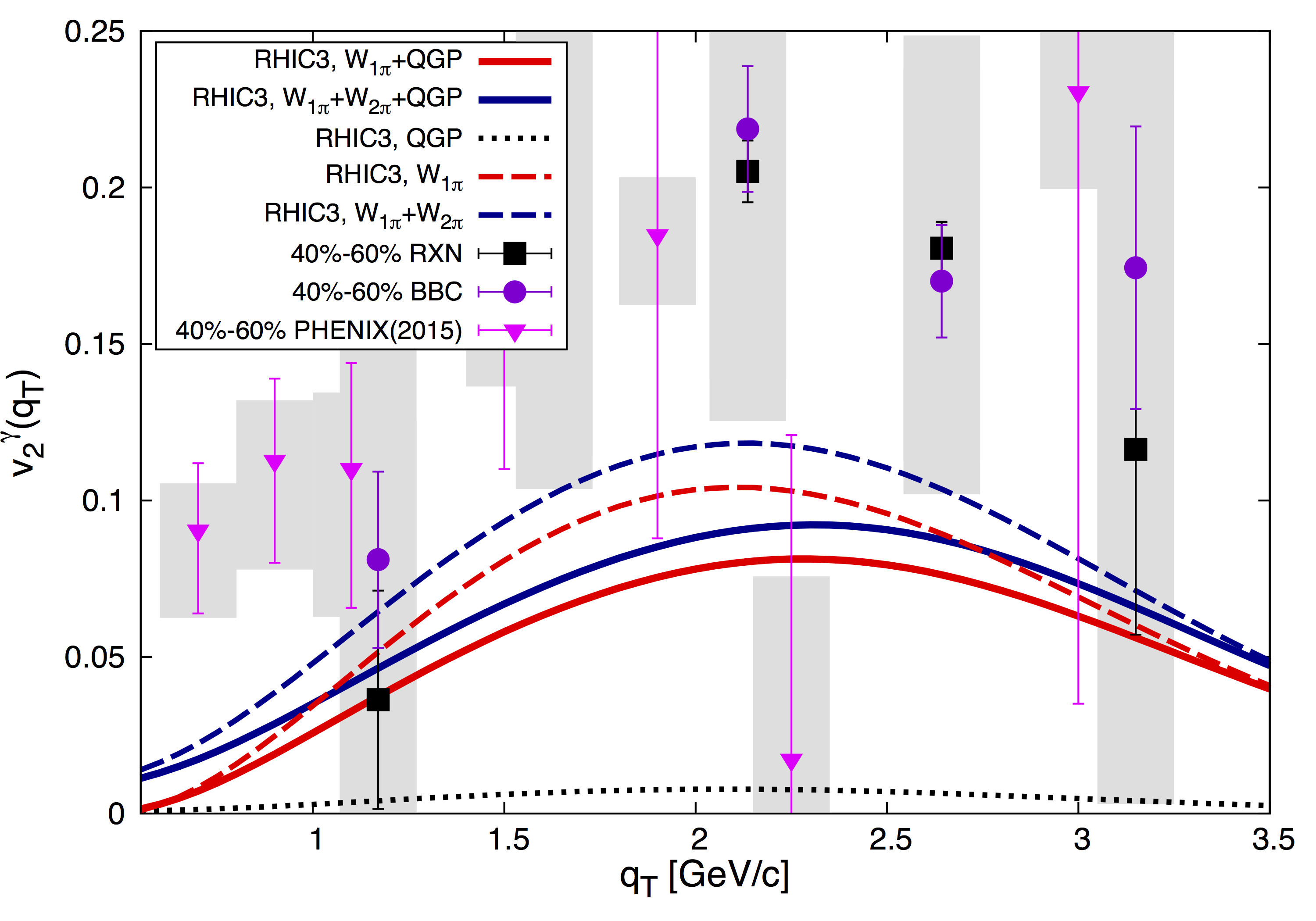}}

\caption{(PHENIX: direct photons) 
Direct-photon elliptic flow $v_2^\gamma$ for  PHENIX~\cite{PHE12,PHE16}.
For the hadronic part before the phase transition, 1st order (CRF1 in red color; up to one pion contribution) and 2nd order (CRF2 in blue color;  up to two pion contribution) hadronic resonance gas (HRG) model  is used. In this calculation, the critical temperature ($\rm T_{crit}$) is 190 MeV and the freeze out temperature ($\rm T_{FO}$) is 137.44 MeV.  
The solid lines are full simulation results from $\rm T_{init}$ to $\rm T_{FO}$ with the phase transition from QGP to HRG. The dashed and dotted lines are the results with partial contributions from the  HRG (below $T_{\rm crit}$) and the QGP (above $T_{\rm crit}$), respectively. {The vertical error bars on each data point indicate the statistical uncertainties and the grey shaded regions indicate the systematic uncertainties of the experiment.}
}
\label{fig2}
\end{figure}

In Fig.~\ref{fig2},  the elliptic flow ($v_2$) of direct-photons  for PHENIX are summarized~\cite{PHE03a,Adl03,Adl04,PHE12,PHE16}. 
The hadronic contributions are referred to as one-pion and two-pion  following from the chiral reduction formulae (\ref{WWW}).
The solid lines are the full simulation results from $\rm T_{init}$ to $\rm T_{FO}$. The dashed and dotted lines are the results with partial contribution from the hadrons below $T_{\rm crit}$ and the AMY resummed 
QGP above $T_{\rm crit}$. The centrality ranges shown below the figures are based on the experimental data. In Fig.~\ref{fig2}, we see that the difference between the one-pion (red) and two-pion 
(blue) contributions to the flow are not significant. Since the prompt yield following from  (\ref{ap-eq:5}) becomes comparable to our hadronic yield  at small $p_T$, it adds to the depletion of the photon flow in this range.
Overall, our elliptic flow for the photons is lower than the one reported  by PHENIX~\cite{PHE03a,Adl03,Adl04,PHE12,PHE16}.

\subsection{Elliptic Flow at LHC}

\begin{figure}[b]
\centering

\subfigure[~0-10\% centrality]{\includegraphics[scale=0.4]{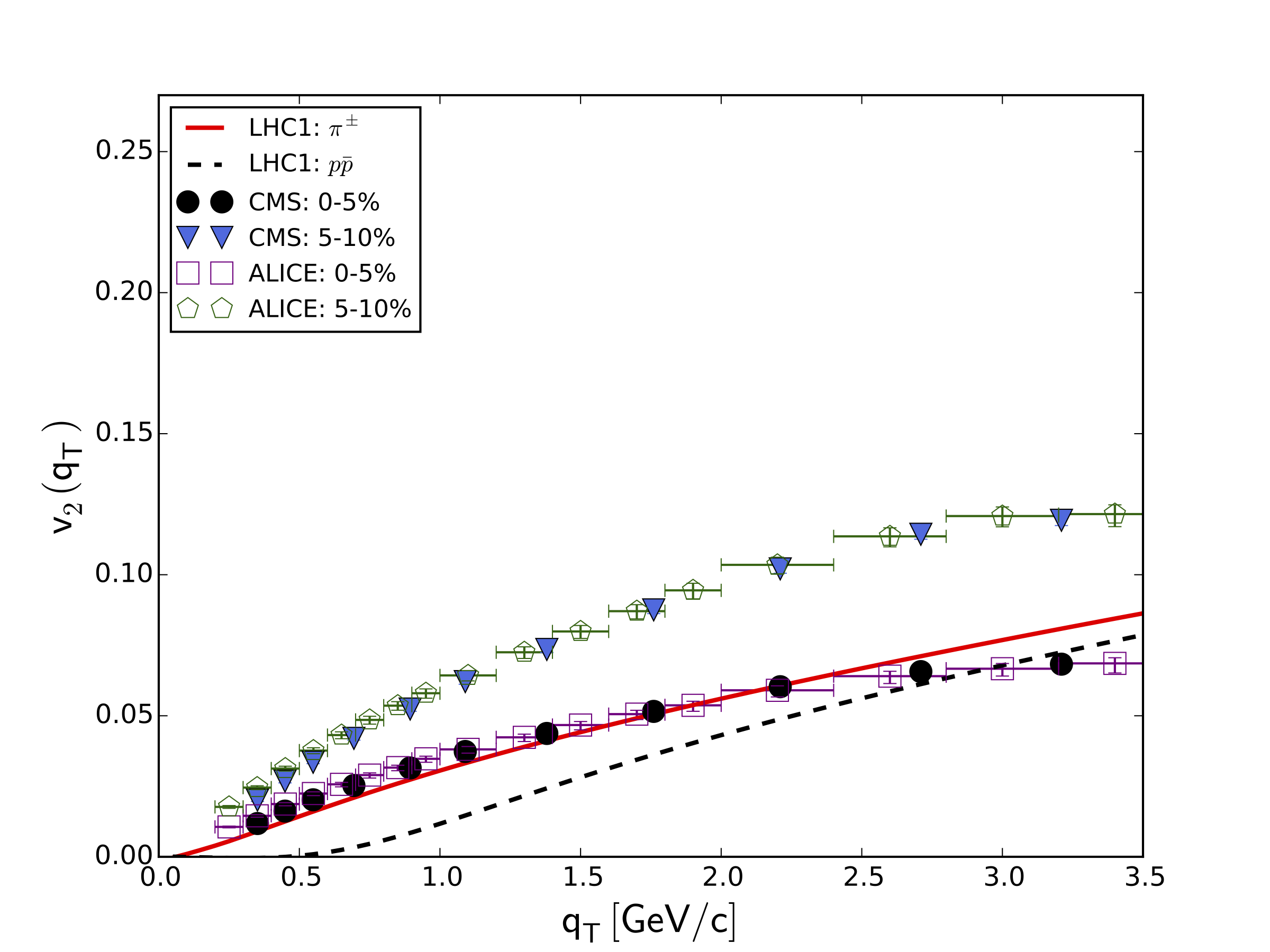}}
\subfigure[~10-20\% centrality]{\includegraphics[scale=0.4]{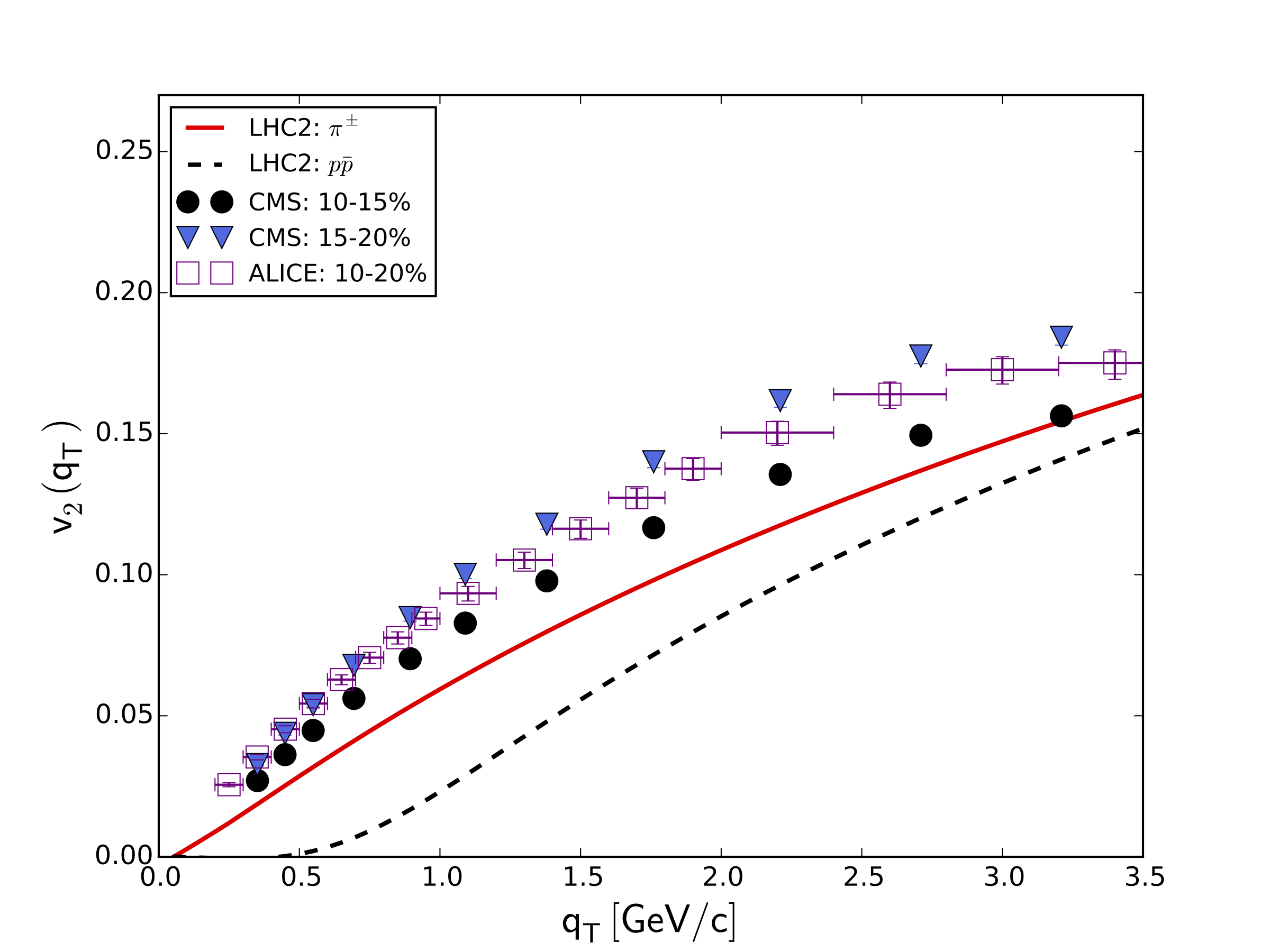}}
\subfigure[~20-30\% centrality]{\includegraphics[scale=0.4]{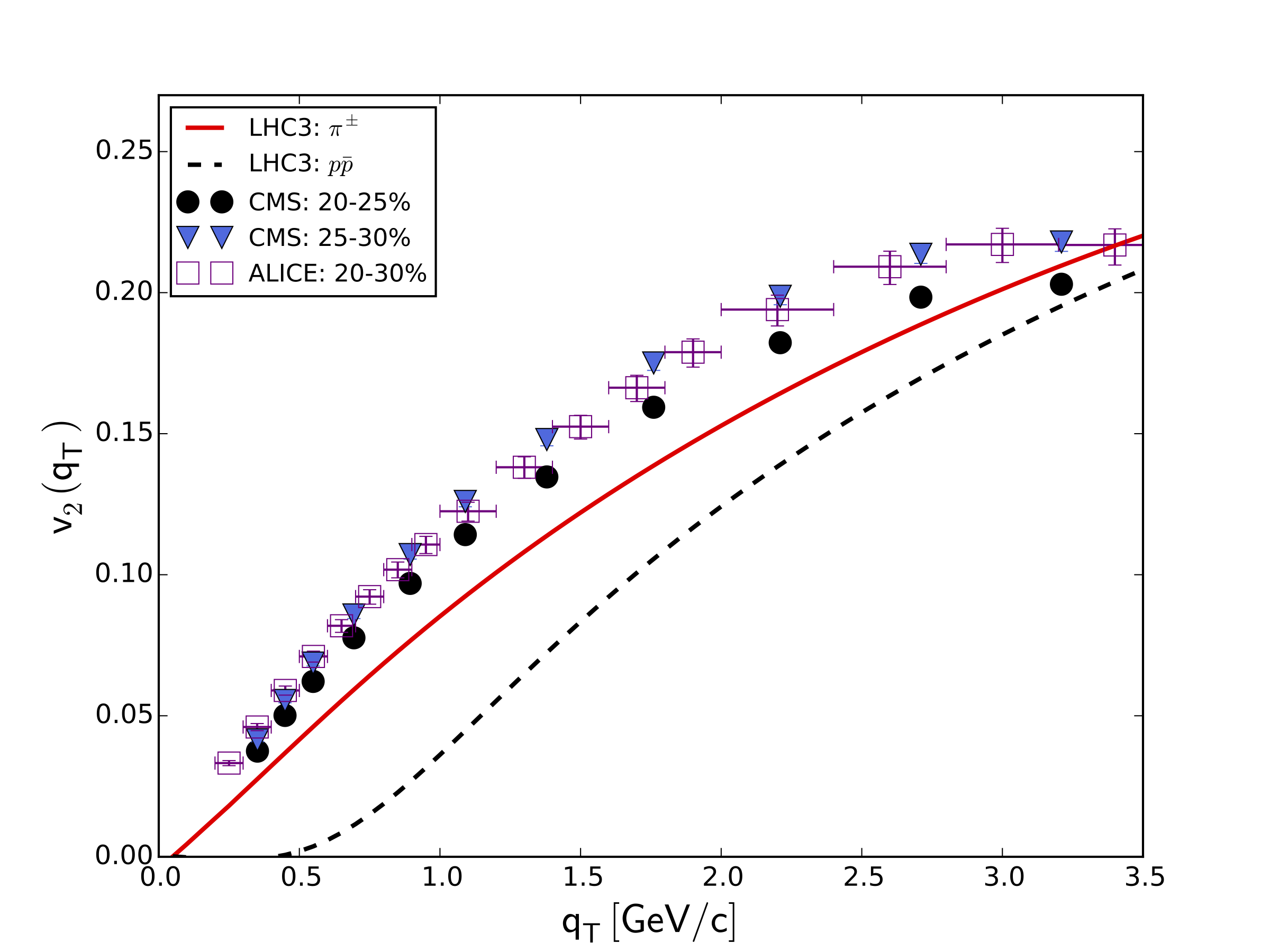}}
\subfigure[~30-40\% centrality]{\includegraphics[scale=0.4]{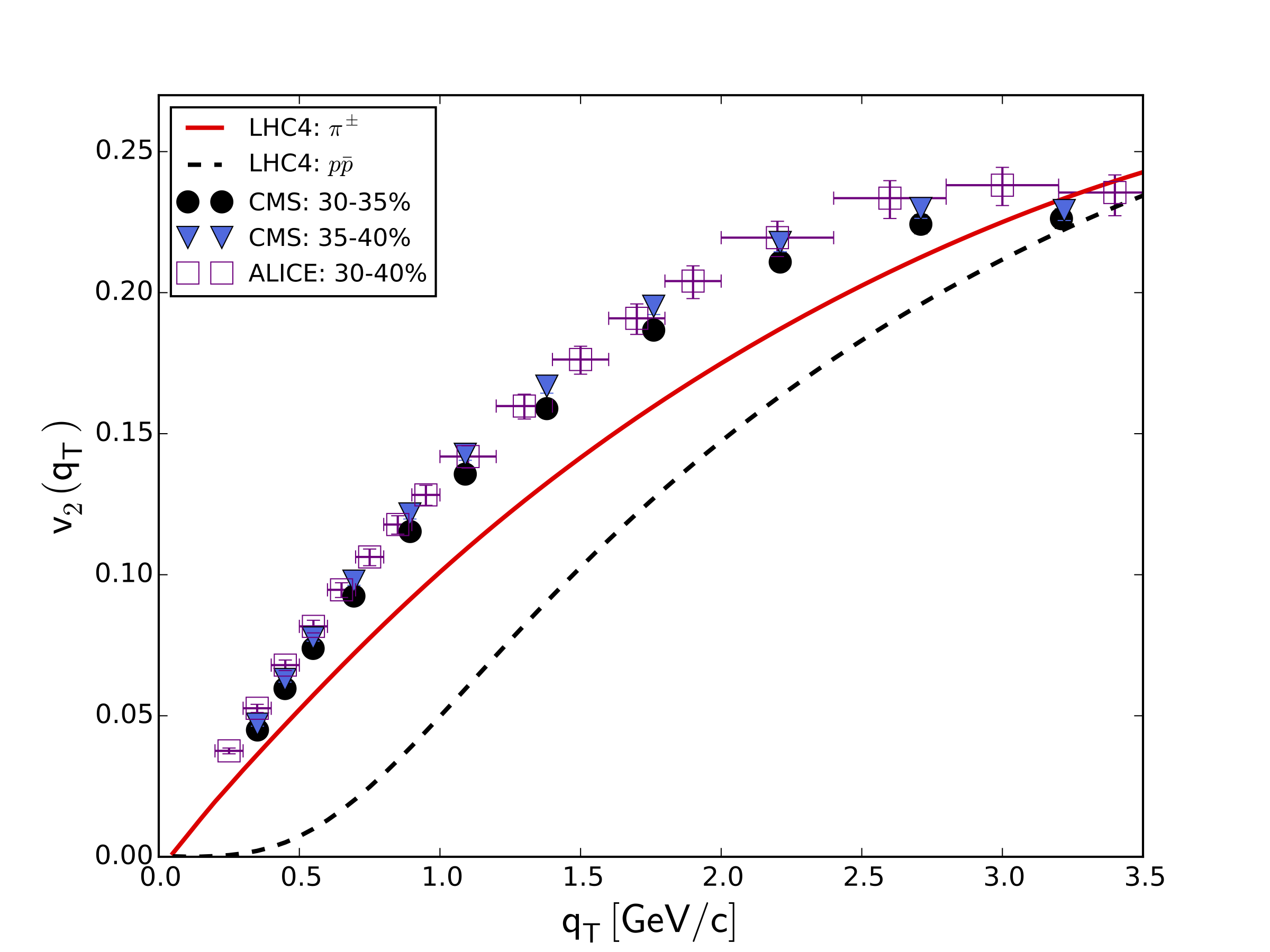}}
 
\caption{(ALICE/CMS: charged particles) Elliptic flow $v_2$ of charged particles ($\pi^\pm, p, \bar p$) for ALICE and CMS~\cite{ALICE13,CMS13}. The parameters for the simulations are summarised in Table~\ref{tab2}. We show the best-fit results in a given centrality region. 
}
\label{fig3}
\end{figure}

In this section, we summarise our best-fit simulation results for the
photon elliptic flow for   ALICE/CMS (\textsuperscript{208}Pb+Pb at $\sqrt{s_{NN}}=2.76$ TeV). 
In Fig.~\ref{fig3}, the simulation results for ALICE/CMS $v_2$ of charged particles are summarised and compared with  the
experimental results~\cite{ALICE,ALICE13,CMS11,CMS13,CMS14}. The parameters used in our simulations are summarised in Table~\ref{tab2}. We used the nucleon-nucleon inelastic cross section $\sigma_{inel}^{NN} = 64.0$ mb that is consistent  with the referred cross section for the Glauber model,  $\sigma_{inel}^{NN} = 64\pm 5$~\cite{CMS14}.   
Note that the numbers of produced charged particles ($N_\pi$ and $N_p$) in our simulations are smaller than those in experiments because our simulations do not account for resonance decays.
The simulated elliptic flows of charged particles for ALICE/CMS are below the empirical values except for the low centrality 0-5\% experimental data.
In Fig.~\ref{fig3}, as the centrality increases, in general, the elliptic flow $v_2$ increases for both the
simulation and experiment.

\begin{figure}[b]
\centering

\subfigure{\includegraphics[scale=0.6]{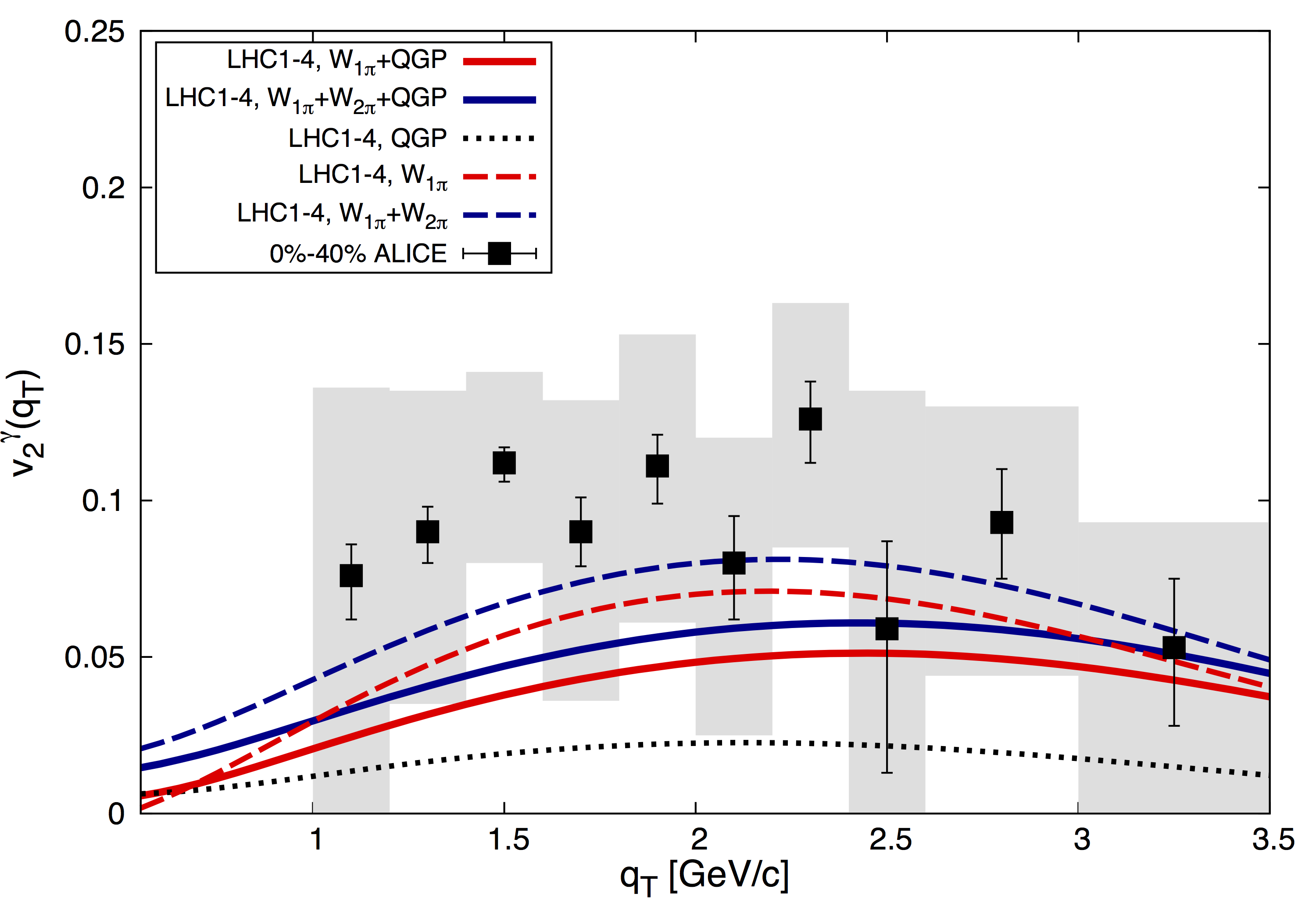}}

\caption{(ALICE: direct photons) 
Direct-photon elliptic flow $v_2^\gamma$ for ALICE~\cite{ALICE12}. 
Before the phase transition, 1st order HRG (CRF1 in red color; upto one pion contribution) and 2nd order HRG (CRF2 in blue color; upto two pion contribution) are used. 
In this calculation, the critical temperature ($\rm T_{crit}$) is 190 MeV and freeze out temperature ($\rm T_{FO}$) is 131 MeV. The initial temperatures ($\rm T_{init}$)  are summarized in Table~\ref{tab2}. 
The solid lines are full simulation results from $\rm T_{init}$ to $\rm T_{FO}$. The dashed and dotted lines are the results with partial contributions from the HRG (below $T_{\rm crit}$) and the QGP (above $T_{\rm crit}$), respectively. Note that the centrality range of the experimental data are 0-40\% . LHC1-4 is an average of our 4 predictions over 4 centrality ranges weighted by the number of photons at each centrality range~\cite{Lohner13}. {The vertical error bars on each data point indicate the statistical uncertainties and the grey shaded regions indicate the systematic uncertainties of the experiment.}
}
\label{fig4}
\end{figure}

In Fig.~\ref{fig4},   the elliptic flow ($v_2$) of direct-photons for ALICE are summarized~\cite{ALICE,ALICE13,CMS11,CMS13,CMS14,ALICE12}\footnote{Experimental data of direct-photons for ALICE is digitized from Fig.5 in \cite{ALICE12} using a graph digitizer software, GraphClick (http://www.arizona-software.ch/graphclick/) }.
Before the phase change from partonic to hadronic, both 1st order (upto one pion contribution) and 2nd order (upto two pion contribution) HRG are used.   The solid lines are the full simulation results from $\rm T_{init}$ to $\rm T_{FO}$. The dashed and dotted lines are the results with partial contributions from the HRG below $T_{\rm crit}$ and the QGP above $T_{\rm crit}$.   Note that the centrality ranges of the experimental data are only 0-40\% in all plots in Fig.~\ref{fig4}. 
Overall, our photon $v_2$ is again smaller than the one reported by ALICE.

\section{Conclusions}
\label{sec:con}

We have assessed  the  elliptic flow of charged particles and direct photons at RHIC and LHC using the 
recently revisited electromagnetic current-current correlators~\cite{Lee14}. 
In order to determine the physical parameters of the hydrodynamic simulations for given centralities, we used the experimentally
reported  multiplicities of charged particles~\cite{Adl04,Adl03,ALICE13,CMS13,ALICE}.  We have estimated the elliptic flow of charged particles and direct photons. For the electromagnetic radiation, we used  the HRG rates  for ($T_{\rm crit}$) and the resummed
AMY QGP rates above $T_{\rm crit}$.  

 For PHENIX, our simulated elliptic flows of charged particles are somewhat below the experimental data in most of the ranges of transverse momenta, especially at high transverse momentum as in Fig.~\ref{fig1}. 
  In contrast, for ALICE and CMS, the simulated elliptic flows of charged particles are below the empirical values at low transverse momentum but reach the empirical values at high transverse momentum $\sim 3.5$ GeV/c as in Fig.~\ref{fig2}.
For the elliptic flow of direct photons, the situation is more subtle because the observed direct photons can be generated in both 
the HRG and the QGP phase.  Overall, we found that our photon flows are lower than the ones reported by both the
PHENIX and ALICE collaborations. 


Our  direct photon yields are larger than those reported in~\cite{PAQUETX}  for the range of
$p_T$ and centralities discussed above, due in part to our larger AMY rates when using a running 
strong coupling $\alpha_s$. However, our photon flows are somehow smaller. There are several
reason for this: 1/  our AMY rates are larger, which means more early photons with lower $v_2$;
2/ our hadronic $v_2$ are slightly lower than those reported empirically, dragging our photon $v_2$;
3/ our freeze-out temperature for the photons is higher than the one used in~\cite{PAQUETX}, to allow 
for extra photon emission from late stage  resonance decays.

Finally, we note that our photon rates calculations do not include the subtle effects of the viscous corrections as
discussed in~\cite{VISCOR}. The chief reason is that our chiral reduction scheme is not reducible to specific diagrammatic
contributions for which local distribution functions with viscous corrections can be implemented. It is an open problem on how
to implement the effects of viscosity in the chiral reduction scheme. Short of this, we cannot reliability assess the contribution
of these missing effects on our final results.

\section*{Acknowledgements}

We thank Kevin Dusling for discussions. We particularly thank Jean-Francois Paquet for making his prompt $pp$
spectra available for comparison and many discussions. 
The work of YMK and CHL was supported by the 
National Research Foundation of Korea (NRF) grant funded by 
the Korea government (MSIP) (No. 2015R1A2A2A01004238 and No. 2016R1A5A1013277).
The work of DT and IZ was supported in part by US DOE grants DE-FG02-88ER40388
and DE-FG03-97ER4014.

\pagebreak



\end{document}